\documentclass{emulateapj}
\usepackage{pslatex}
\def\deg{{\rm o}}

\def\idm#1{{\mbox{\scriptsize #1}}}

\newcommand\Chi{{(\chi^2_\nu)^{1/2}}}
\def\astrobj#1{#1\ }

\def\url#1{\texttt{#1}}
\newcommand\stara{{\astrobj{HD~128311}}}
\newcommand\starb{{\astrobj{HD~82943}}}

\begin{document}
%
%
%
\title{Trojan pairs in the HD~128311 and HD~82943 planetary systems?}

\author{Krzysztof Go\'zdziewski\altaffilmark{1}}
\affil{Toru\'n Centre for Astronomy, N.~Copernicus University,
Gagarina 11, 87-100 Toru\'n, Poland}
\author{Maciej Konacki\altaffilmark{2}}
\affil{
Nicolaus Copernicus Astronomical Center, Polish Academy of Sciences,
Rabia\'nska 8, 87-100 Toru\'n, Poland\\
Department of Geological and Planetary Sciences, California
Institute of Technology, MS 150-21, Pasadena, CA 91125, USA
}
 
\altaffiltext{1}{e-mail: k.gozdziewski@astri.uni.torun.pl}
\altaffiltext{2}{e-mail: maciej@ncac.torun.pl}

\begin{abstract}
Two nearby stars, HD~128311 and HD~82943, are believed to host pairs of
Jupiter-like planets involved in a strong first order 2:1 mean motion resonance
(MMR). In this work we reanalyze available radial velocity (RV)  measurements
and demonstrate that it is also possible to explain the  observed RV variations
of the parents stars as being induced by a pair of Trojan planets (i.e., in a
1:1 MMR).  We show that these Trojan configurations reside in extended  zones of
stability  in which such systems may easily survive in spite  of large masses of
the planets,  large eccentricities and nonzero mutual  inclinations of their
orbits. We also show  that HD~82943 could harbor a previously unknown third
planet of $\sim$0.5  Jupiter masses in $\sim 2$~AU orbit.
\end{abstract}


%
%
\section{Introduction}
%
%

Follow-up radial velocity (RV) observations of Sun-like stars with planets have 
revealed a number of extrasolar multi-planet systems. Many of them are  involved
in low-order mean motion resonances (MMRs). In particular, at least four
extrasolar systems are involved in a strong first order 2:1~MMR:   Gliese~876
\citep{Marcy2001}, HD~82943 \citep{Mayor2004}, HD~128311~\citep{Vogt2005}, and
HD~73526~\citep{Tinney2006}. A considerable effort has been devoted to study the
origin \citep{Kley2003,Kley2004}  and dynamical stability of these intriguing
systems 
\cite[e.g.,][]{Gozdziewski2001b,Lee2002,Ji2003,Beauge2003,Lee2004,Psychoyos2005,
FerrazMello2005,
Lee2006}. Yet dynamical studies of the resonant configurations often rely on
the  2-Keplerian coplanar fits by the discovery teams. It has been demonstrated
that the 2-Keplerian models can be of a very limited use  for systems involved
in strong mutual interactions \cite[e.g.,][]{Laughlin2001,Rivera2001,Gozdziewski2005a}. 
Even if a model  incorporates mutual interactions, due to
typically  short time-span and a limited number of the observations, the
orbital  inclinations are barely constrained and usually only coplanar, edge-on
configurations  are considered. However, the recent results of
\cite{Thommes2003} and  \cite{Adams2003} suggest that a significant fraction of
planetary systems  involving giant planets may be substantially non-coplanar.
Dynamical  mechanisms which lead to fast amplification of the relative
inclination are  especially effective in the first order resonance
configurations \citep{Thommes2003}.  Also dynamical relaxation and collisional
scattering of the protoplanets may  favor large relative inclinations in such
systems, even if they initially emerge in a flat protoplanetary disk. 

The interpretation of the RV data for multi-planet system may be difficult. 
The determination of the number of planets and their orbital periods  can be
problematic for some systems. In particular, for those involved in a 1:1~MMR.  
A periodogram of  the RV signal for such a system \citep{Laughlin2002} is 
basically indistinguishable from that of a single planet in an eccentric orbit  
or, as we show in this paper, from a periodogram of a 2:1~MMR orbital configuration. 
\cite{Laughlin2002} and \cite{Nauenberg2002} have demonstrated that a coplanar
1:1 MMR of Jovian planets may be stable in a wide range of their orbital
parameters. The results of hydrodynamic simulations by \cite{Laughlin2002} 
indicate that Trojan planets in tadpole or horseshoe orbits might readily form 
and migrate within a proto-planetary disk.  Presumably, a 1:1 configuration may 
also emerge as a result of dynamical relaxation or migration frequently used to 
explain 2:1 MMR configurations. In the Solar System  there exist a number of
moons  involved in this type of resonance --- famous Janus-Epimetheus system 
(co-orbital moons of Saturn, exchanging orbits), Helene-Polydeuces (Trojans of
Dione, a moon of Saturn), Telesto--Calypso (Trojan  moons of Tethys, yet another
moon of Saturn).  Dynamically, these configurations mimic planetary systems in
the 1:1~MMR.

In this paper, we perform an independent analysis of the RV data for  \stara and
\starb to verify if the observed RV variations can be  explained not only by a
configuration in a 2:1~MMR but also 1:1~MMR. The inevitable problem with
modeling such systems is that due to a limited number of data and relatively
large measurement errors, the  best-fit orbital elements often and easily lead
to catastrophically  unstable configurations. In order to solve this problem one
needs a method  of fitting which incorporates a stability criterion. Without
such a  constraint, one can find a stable best-fit orbits basically by chance.
When we deal with a 1:1 MMR configuration, an appropriate stability  control is
essential as the planets share the same (or a very similar)  orbit and a
multi-parameter dynamical model is highly nonlinear. In this paper we use the 
term ``Trojan planets'' not only for tadpole, close to coplanar and circular 
configurations but for all configurations characterized by a 1:1 MMR so  having
not only similar semi-major axes but also (possibly) large relative 
inclinations and variable eccentricities.

\section{Numerical approach}

Due to strong mutual interactions, the planetary systems with giant planets 
have to avoid the unstable zones of the MMRs, a proximity of the collision  zone
and the zone of global instability where the overlapping of MMRs occurs. 
Otherwise, the chaotic diffusion quickly leads the planets to collisions
between  each other or with the parent star. The overall picture of the phase
space of a planetary system is predicted by the fundamental
Kolmogorov-Arnold-Moser  theorem \citep{Arnold1978}: the phase space is {\rm not
continuous} with respect to the stability criterion.  Hence, commonly used (in
particular, gradient-like) algorithms of exploring  the phase space are poorly
designed for this task because they are "blind" to a sophisticated fractal-like
structure of the phase space. 

The KAM stability is described in terms of stable (regular, quasi-periodic) and
unstable (chaotic)  motions.  At first, the use of such a formal criterion in
the fitting process  may be problematic. Almost any planetary system, including
our own, can be very close to a chaotic state. Nevertheless, we expect that
even  if chaos appears, it should not impair the astronomical stability
\citep{Lissauer1999}  meaning that a system is bounded over a very long time and
any collisions or ejections  of planets do not occur. However, for
configurations involving Jupiter-like companions in close orbits with large
eccentricities, the formal stability seems to be well  related to the
astronomical stability of the system, i.e., chaotic motions mean a fast 
destabilization of a planetary configuration over short-time scale related to
the  most significant, low-order MMRs.   It has been already demonstrated by
dynamical analysis of systems residing in the regions of the phase space where
the low-order MMRs are possible
\cite[e.g.][]{Gozdziewski2001b,Gozdziewski2005a,Gozdziewski2006}. We should
note that  there is not known any general relation between the Lyapunov time (a
characteristic time-scale of the formal instability) and the event time
\cite[the time after which a physical change of a planetary system happens; see,
e.g.][]{Lecar2001,Michtchenko2001}.  To put these ideas into  action and to
search for stable best-fit solutions in a self-consistent  and optimal fashion,
we treat the dynamical behavior (in terms of chaotic  and regular or
mildly-chaotic states) as an additional observable at the same  level of
importance as the RV measurements. This powerful approach has been  already
described in detail and successfully applied in 
\cite{Gozdziewski2003e,Gozdziewski2005a,Gozdziewski2006}.

The kernel of our approach is the the genetic algorithm scheme (GA) implemented
by \cite{Charbonneau1995} in his publicly
available\footnote{\url{http://www.hao.ucar.edu/Public/models/pikaia/pikaia.html}}
code PIKAIA. The GAs are ideal for our purpose because of their global
non-gradient nature and their proven ability to efficiently explore a
multidimensional noncontinuous parameter space. The RV data are modeled by a
synthetic signal of the full $N$-body dynamics~\citep{Laughlin2001}. The $\Chi$
function is modified by a stability penalty term employing an efficient fast
indicator MEGNO \citep[]{Gozdziewski2003e}.  The GA fits are finally refined by
yet another very accurate non-gradient minimization scheme by Nelder and Mead
\citep{Press1992} widely known as the simplex method. This  greatly reduces the
CPU usage. We give the algorithm an acronym GAMP (Genetic Algorithm with MEGNO
Penalty).

The simplex method finds local minima of $\Chi$. The code may be
also trapped in resonance islands surrounded by strongly chaotic motions even if
$\Chi$ inside such islands is larger than in the neighboring  (but unstable)
areas. By collecting solutions to which the GAMP converged in many independent
runs, we gather  an ensemble of the local best-fit solutions. It helps us to
illustrate the multidimensional  properties of $\Chi$ and to obtain realistic
estimates of the parameter's errors  by choosing the solutions within prescribed
limits of the overall best-fit $\Chi$ found  in the entire search. At the end, some of
the selected best fits can be refined  with longer integration times and much
lower simplex tolerance than is used during the search phase.

Finally, the stability of the best-fit solutions is examined in planes of
selected orbital osculating elements using the Spectral Number method ($SN$)  by
\cite{Michtchenko2001}. It is an efficient fast indicator completely independent
on MEGNO. This enables us to verify and illustrate the best-fit solutions in a
robust way and to examine dynamical properties of such configurations in wide
ranges of neighboring initial conditions. Note that the $SN$ is related here to
the short-term dynamics. Thus the spectral signal analyzed is a time series  $\{
f(t) = a(t) \exp [\mbox{i} \lambda(t)]\}$ where $a(t)$ and $\lambda(t)$  are
respectively an osculating {\em canonical} semi-major axis and longitude of a
planet. Such an analysis makes it possible to  resolve the proper mean motion
$n$ as one of the fundamental frequencies of the system. Note also that a
stability criterion in the GAMP code may be basically arbitrary.  We use the
formal KAM criterion as the most general and well defined one.

\section{\stara}

The HD~128311 is an active K0 star \citep{Vogt2005}. In the discovery paper, 
\cite{Butler2003} found an indication of a Jovian planet and a linear trend in
the RV data. They concluded that due to photospheric activity ($\log
R_{\idm{HK}}' \simeq -4.4$)  the stellar jitter is large $\sim 20$~m/s and the
signal variability may be explained exclusively by the jitter. Using an updated
set of 76 RV measurements, \cite{Vogt2005} found that the observations can be
modeled by a system  of two Jupiter-like planets involved in a 2:1 MMR. The
current estimate of the stellar jitter by these authors is $\sim 9$~m/s but
still uncertain  with a $50\%$ error. We rescale the measurement errors by
adding this estimate in quadrature to the formal RV errors.  

The discovery  team
reports that the best-fit 2-Keplerian model   yielding $\Chi =1.86$ and an rms
$\sim 18$~m/s is catastrophically unstable. 
Using our hybrid GA/simplex code \citep{Gozdziewski2006a} driven by the 
Keplerian model of the RV, we found a different, apparently better 2-planet
solution which has $\Chi=1.717$ and an rms $=15.16$~m/s. The model parameters 
$(K,P,e,\omega,T_{\idm{p}}-T_0)$, i.e., the semi-amplitude, orbital period,
eccentricity, the argument of periastron and the time of periastron passage, for
this fit are as follows: 
(51.948~m/s,    459.870~d,      0.362,     59.401$^{\circ}$,   2474.867~d) and
(77.214~m/s,    917.371~d,      0.248,      5.541$^{\circ}$,   2310.806~d) 
for the inner and outer planet, respectively; $T_0=$ JD~2,450,000 and the
velocity offset $V_0 = 1.011$~m/s. It is argued that the fit parameters of
a multi-planet system should be interpreted in terms of osculating Keplerian
elements and  minimal masses related to the Jacobi coordinates \citep{Lee2003}.
Adopting the date of the first observation as the osculating epoch, we
recalculated the inferred {\em  astrocentric} osculating elements   ($m_p \sin
i,a,e,\omega,M$) as
follows 
(1.639~\mbox{m$_J$}, 1.101~AU, 0.362, 59.37$^{\circ}$, 272.79$^{\circ}$)  and 
(3.194~\mbox{m$_J$}, 1.746~AU, 0.249, 5.39$^{\circ}$, 199.62$^{\circ}$) 
for the inner and outer planet, respectively.  Still, the
derived configuration is also unstable and disrupts during about 200,000~yr. 
Nevertheless, we found that its MEGNO signature is characteristic for a system
residing on the border of a stable region rather than a collisional configuration.
Thus one may suspect that in its proximity, a rigorously stable solutions can be
easily found.

In order to deal with the problem of an unstable 2-Kepler fit, \cite{Vogt2005} 
applied a method of fitting which incorporates the mutual interaction between
planets \citep{Laughlin2001}  and, additionally, explicitly involves stability
criterion. As such the authors use the maximal eccentricity attained by the
companions during an integration time. They report many stable solutions
corresponding to the 2:1 MMR.  According to the authors, their best fit yields
an rms $\sim 14.7$~m/s and $\Chi\sim 1$. We could not reproduce that value of
$\Chi$. The quoted $\Chi$ could be misprinted or a jitter estimate larger than
$\sim 9$~m/s was used in the calculations. 

For a comparison with that result and as a background for a further analysis,  
we performed the GAMP search for the best-fit solution to the RV data  from
\cite{Vogt2005} assuming that a 2:1 resonance is indeed present in the coplanar
and edge-on system. We also explored a more general model in which the orbits  
are mutually inclined but we did not find substantially better fits.

In Fig.~\ref{fig:fig1}, the elements of the best-fit solutions from the  GAMP
runs are shown in a few representative planes of the osculating  elements at the
date of the first observation, JD, 2,450,983.827.  In the GAMP code, the MEGNO
was evaluated over $1000-5000$ orbital periods of the outer planet that is both
efficient enough and makes it possible to withdraw strongly chaotic, unstable
solutions. The best-fit initial conditions are marked with symbols of different
sizes ---  larger circles indicate a smaller $\Chi$ (a better fit). Only {\em
stable}  solutions within the $3\sigma$ confidence interval of the best-fit
solution (given in Table~1) are shown. In overall, the statistics of initial
conditions shown in Fig.~\ref{fig:fig1}  is in accord with the results of
\cite{Vogt2005}, see their Fig.~12. The permitted 
{\em initial} eccentricities of the fits
span skewed and narrow valley in the $(e_{\idm{b}},e_{\idm{c}}$)-plane. Let us
note that we represent the $N$-body initial conditions in terms of astrocentric,
osculating Kepler elements at the epoch of the first observation.

The orbital elements for the outer planet have  larger errors than for the inner
one. Both semimajor-axes and  phases are already very well determined. The
parameters of the stable best-fit solution are given in Table 1 (Fit I). Note
that this strictly regular solution has $\Chi \simeq 1.731$  and rms $\simeq
15.28$~m/s. A very similar value of $\Chi$ in the $N$-body and Keplerian fits
means that the mutual interactions between planetary companions are not evident
in the Doppler signal spanning  10~yr. Nevertheless, we stress that the
stability constraints are essential for obtaining stable 
configuration of the $N$-body
model. 

In Figure~\ref{fig:fig2}, we show the stability analysis the
best fit  solution
corresponding to a 2:1 MMR (Fit~I given in Table~1). The
Spectral Number, $\log SN$, as well as $\max e_{\idm{b,c}}$, the maximal
eccentricities of both planets,  and $\max \theta$, the maximal
$\theta=\varpi_{\idm{b}}-\varpi_{\idm{c}}$ (where $\varpi_{\idm{b,c}}$ are the
longitudes of pericenters), attained during the integration over $\sim
7\cdot10^4$
orbital periods of the outer body are shown in Figure~\ref{fig:fig2}. It turns
out that the best-fit solution lies on the border of an island related to the
corotation of apsides ($\theta$ as well as the
critical arguments of the 2:1 MMR are librating about $0^{\circ}$ with a
large amplitude). The border of the  stable resonance zone which is present in
the $SN$-map can be also seen in all  other maps, in particular in the $\max
e$-maps. This is a strong argument that the formal stability criterion is in the
one-to-one relationship with  the behavior of the system. Clearly, the search
zone for the best-fit solution should be limited to the resonance island which
constantly changes  its shape when we change the orbital parameters. Thanks to
the instability  penalty in our approach, we have confidence that the obtained 
solution is indeed optimal (i.e., it minimizes $\Chi$ and is 
dynamically stable).

In our next test, we carried out a search for a stable Trojan configuration.
Our  model was extended to 14 osculating orbital elements including the
inclinations and one nodal longitude as free parameters. Note that due to  very
similar orbital semi-major axes, the planets are numbered by giving the
symbol~"b" to that planet which has a smaller initial eccentricity. As one can
see in Fig.~\ref{fig:fig3},  a well defined minimum of $\Chi$ is present in  the
$(a_{\idm{b}},a_{\idm{c}})$ and  $(e_{\idm{b}},e_{\idm{c}})$ planes. The
best-fit inclinations are not very well constrained nevertheless their concentration
is quite evident, in spite of a moderate time span of the observations.

The osculating elements of the best-fit solution are given in Table~1 (Fit II).  
Its $\Chi\simeq 1.797$ and rms $\simeq 15.48$~m/s are very close to those of the
2:1 MMR configuration. The synthetic RV signals of both solutions are shown in
Fig.~\ref{fig:fig4}.  They can be barely distinguished one from another.
We also computed the Lomb-Scargle periodograms of both synthetic  signals
and we plotted them together with the  periodogram of the data set in
Fig.~\ref{fig:fig5}. It shows that periodograms of the 2:1 and 1:1
configurations almost perfectly match each other. It would be very difficult
to distinguish between the configurations by looking only at the periodograms.

The best-fit Trojan configuration resides in a wide stable zone  in the plane of
the eccentricities which extends up to 1 for both of these  elements (see
Fig.~\ref{fig:fig6}). The resonance area in the 
$(a_{\idm{c}},e_{\idm{c}})$-plane covers about 0.2~AU.  This width is even
larger than for the 2:1 MMR configuration (see Fig.~\ref{fig:fig2}). Obviously,
the stable 1:1 MMR is possible due to the corotation of the  apsides seen in the
$\max \theta$-maps ($\theta$ librates about $180^0$). 

Our choice of the stability criterion enables us to obtain very sharp borders 
of the resonance area. If this criterion was violated, the system would quickly 
disrupt (because both $e$ go to 1). In Fig.~\ref{fig:fig7} we show the
stability  maps for a solution which has an rms of about $15.7$~m/s (slightly more
than the best one) and corresponds to much smaller masses (both initial
inclinations are about  $45^{\deg}$). Essentially, all the dynamical features of
the system do not change  but the width of the resonance zone shrinks
substantially. This is also an argument  that an appropriate stability criterion
has to be an integral part of the fitting  tool. Due to an extremely nonlinear
nature of the system, even a small change of its initial elements may lead to a
significant change in the shape of the resonance zone. Simultaneously, $\max e$
becomes very "flat" in the regions of  large $e$ --- so $\max e$ would not be a
convenient stability indicator in  a GAMP-like code. Another argument is that
the variable rate of the chaotic diffusion which leads to the changes of the
eccentricity can be sometimes to small to detect a collision or a qualitative 
change of the configuration over relatively short integrations which, for
efficiency reasons, have to be limited to the time-scale of the MMRs. One might
think that (again, mainly for efficiency reasons) $\max \theta$  would be a
better choice than both $\max e$ and MEGNO as a stability criterion. However,
that test may also fail as in the resonance  zone the critical
angle $\theta$ may librate about different centers (usually, about of
$0^{\circ}$ or $180^{\circ}$) but it can  circulate in some marginally unstable
regions as well (see Appendix for details).

\section{\starb}

The \starb planetary system \citep{Mayor2004} has drawn attention of  many
researchers \cite[e.g.,][]{Gozdziewski2001b,Ji2003a,FerrazMello2005,Lee2006}. 
The 2-Keplerian solutions produced by the discovery team correspond to  
catastrophically unstable configurations
\citep{Gozdziewski2001b,FerrazMello2005,Lee2006}.  The discovery team did not
publish the observations of HD~82943 in source form. The method of dealing with the problem
of unavailable RV data relies on  digitizing the published figures depicting the
measurements. This somewhat  unusual approach has already become an accepted
procedure 
\citep[e.g.,][]{Gozdziewski2001b,Gozdziewski2004,FerrazMello2005,Lee2006}.

First, we digitized  142 data points from the figures of \cite{Mayor2004}.
They slightly  differ from the real observations. In particular, it is difficult
to  recover the exact moments of the observations \citep{FerrazMello2005} but 
such digitized measurements still properly describe  the overall shape of the
observed RV curve and its characteristic features. We also graphically derived 
the measurement errors and rescaled them by adding in quadrature the stellar 
jitter which we estimated as $\sim 5$~m/s on the basis of \citep{Wright2005}.

Having such "measurements", we recovered the best-fit 2-Kepler solutions
published by the discovery team \citep{Mayor2004}. The only problem seems to be
a little larger value of the rms of $\simeq 7.1$~m/s (compared to  6.8~m/s
quoted in the original work). Using the digitized RV measurements, 
\cite{FerrazMello2005} showed that stable 2:1~MMR configurations are possible. 
Their orbital parameters of coplanar,
edge-on systems are similar to those we found with GAMP (Fit III
in Table~1). We note that these authors looked for the best-fit solutions by 
minimizing the rms rather than $\Chi$, and they did not increased the internal  
errors by jitter. Also the discovery team did not account for the jitter in 
their solutions.

We extended the search for the best-fit solution assuming that a 1:1~MMR can be
present in the \starb system. As in the previous case, we did not find any
stable and strictly coplanar, edge-on configuration of this type. However, using
the generalized model in which masses, inclinations and one nodal  longitude are
free parameters, we found many stable solutions. Their quality is not as good as
for the 2:1~MMR --- the best 1:1 MMR fit has $\Chi \simeq 1.2$ and the rms is
$\simeq 8.1$~m/s which is about of 1~m/s worse than for our best  2:1 MMR
solution. Still, the 1:1 MMR solution may be plausible  (note that we use
digitized "observations"). Also, since the mass of the parent star cannot be
determined precisely \citep{FerrazMello2005},  and when new measurements are 
available, the best-fit parameters and their $\Chi$ may change.  We demonstrate
it in the next section.

The best-fit 1:1 MMR configuration (Fit IV in Table~1) is characterized by initially 
large mutual orbital inclination because,  although the inclinations for both
planets are almost the same,  the nodal longitude is about $180^{\circ}$ (and
the apsidal lines are anti-aligned).  The orbital evolution leads to quite large
variations of the orbital inclinations (a few tens of degrees). The stability
maps shown in Figure~\ref{fig:fig8} reveal that, apparently, such a system would
be locked in an extremely large zone of stable motions which extends up to
$e_{\idm{b,c}} \sim 1$. 
This means that the eccentricities could reach extremely
large values but the system would be still stable. 
The width of the resonance
with respect to $a_{\idm{c}}$ is also relatively large, about $0.2$~AU.  A zone
of strictly periodic motions can be seen in the map for $\max \theta$, close to
its diagonal.

\subsection{Fits for \starb{} revisited}
 
We extended the analysis of \starb after gaining access to the same data set
used by \cite{Lee2006} (and also Lee 2005, private communication). These
authors  also used the ``digitized'' measurements from \cite{Mayor2004} but added
new observations obtained with the Keck/HIRES. These new very accurate
data fill in some gaps in the CORALIE measurements as well as  significantly
increase the time span of the observations. For consistency with 
\cite{Lee2006} we adopted the jitter estimate of $4.2$~m/s.

The results of fitting 2-planet Keplerian models and the analysis of their
stability by \cite{Lee2006} strongly confirm the possibility of a stable 2:1 MMR
in the \starb system. Still, new questions may be asked. The best fit of the 2:1
MMR configuration yields an rms $\sim 8$~m/s which is unexpectedly larger  by
1~m/s from that quoted for the CORALIE data by the discovery team 
\cite{Mayor2004} as well as by \cite{FerrazMello2005}, and in this work. 

For the updated data set, we recovered all the best fit Keplerian solutions
quoted by \cite{Lee2006} using our GA/simplex code. Some of them appear to be formally
chaotic or strongly unstable. Thus we searched for a stable $N$-body solution 
using GAMP assuming that the velocity offsets for the CORALIE and
Keck/HIRES data are independent. The found best-fit, {\em rigorously stable}
solution corresponding to the 2:1 resonance of an edge-on system is given in Table~2 
(Fit~V). Its quality is  not very different from the best solutions found by
\cite{Lee2006} but the initial eccentricities are significantly different.  Our
Fit~V is most similar to the 2-planet Kepler Fit~II of \cite{Lee2006}.
We also found other solutions which are similar to their Fit~III and IV with
respect to small initial $e_{\idm{c}}$. 

According to the results of \cite{FerrazMello2005} and \cite{Lee2006},
$e_{\idm{c}}$ and $\omega_{\idm{c}}$ are the less constrained parameters of 
the Kepler fits to the RV data of HD~82943. Thus we computed dynamical maps 
for the relevant solutions (see Fig.~\ref{fig:fig9} and its caption) as well 
as for Fit~II of \cite{Lee2006}.  These maps reveal two narrow zones of stability 
in which the best-fit solutions reside. All acceptable (stable) 2:1~MMR 
fits likely belong 
to these two distinct islands. We label them with A and B in Fig.~\ref{fig:fig9}.
Note that the positions and shape of the resonance areas are significantly altered 
when the fit parameters are adjusted.  Inside the resonance zone~A, the amplitudes 
of the critical angles may vary in wide ranges. \cite{Lee2006} found that 
their Fit II has very small amplitudes of the critical angles of the 2:1~MMR, 
$\sim 10^{\circ}$. Our best $N$-body fit V yields much larger amplitudes, 
$\sim 40^{\circ}$.  The large amplitudes of the critical angles are also reported by
\cite{FerrazMello2005}. The resonance island~A is characterized by the corotation 
of apsidal lines. 
We found that in this zone $\max \theta$ may be very close to  
the libration center $0^{\circ}$.
For instance, for the dynamical map of Fit~V,
at ($e_{\idm{c}}\sim 0.116,\omega_{\idm{c}}\sim 127.7^{\circ}$) 
the variations of $\max \theta < 2^{\circ}$, indicating a
strictly periodic solution.
In the second island labeled by B in 
Fig.~\ref{fig:fig9}, $\max \theta$ 
also librates about $0^{\circ}$ but with large 
amplitudes.
 
We may speak about dynamical similarity of the best-fit solutions found so
far, having in mind their position in the two resonance zones. The results
of the dynamical analysis done by \cite{Lee2006} and in this paper favor
the 2:1~MMR fits located in the island~A (about $\omega_{\idm{c}} \sim
120^{\circ}$). This zone is extended with respect to the not well constrained
$e_{\idm{c}}$ and very small amplitudes of the 2:1~MMR critical angles are
possible.

We also found many  stable mutually inclined configurations using the orbital
inclinations and one nodal argument as free parameters. Still, all these fits
have the rms $\sim 8$~m/s. By releasing the stability requirements in the GAMP
code, one finds the  best 2-planet fit  yielding the rms $\sim 7.5$~m/s (Fit~VII
in Table~2). However, this configuration disrupts in a few hundred years.
Curiously, such a  solution involves a brown dwarf and a Jovian planet on
inclined orbits (mutual inclination of $\sim 80^{\circ}$).

In fact, our main goal was to perform a possibly extensive search for 1:1
configurations. The statistics of stable solutions gathered in this search is 
illustrated in Fig.~\ref{fig:fig10}, in a similar manner as for \stara system.
Qualitatively, the solutions do not differ from the ones we found using
the CORALIE data only.  The semi-major axes and the eccentricities are very well
constrained. Also the initial inclinations and masses are bounded to two well
determined local minima of $\Chi$. The best fit solution is given in Table~2
(Fit~VI).  Its rms $\sim 8.4$~m/s is even closer to that of the 2:1 MMR best fit
than in the case of the CORALIE data alone. Its MEGNO signature  which 
indicates  perfectly stable, quasi-periodic configuration, and the evolution of
orbital elements during 3~Myr are shown in Fig.~\ref{fig:fig11}. 
The relevant dynamical maps of the best fit in the
($e_{\idm{b}},e_{\idm{c}}$)- and ($a_{\idm{c}},e_{\idm{c}}$)-plane are
shown in Fig.~\ref{fig:fig12}.
We also
compared periodograms (Fig.~\ref{fig:fig13}) of the synthetic curves for the 2:1
MMR, 1:1 MMR and the measurements (shown in Fig.~\ref{fig:fig14}). Also
in this case the periodograms for the 2:1 and 1:1 MMR perfectly match each
other. 

\subsection{The third planet in HD~82943?}

The solutions described so far do not explain the curious rms excess that is
present in the extended data set. It seems unlikely the problem is  caused
by some inconsistency of the measurements from the two spectrographs.  
Another possible explanation is that there is a new, unknown object  in the system.
That possibility is suggested by \cite{Lee2006}. Looking at their Fig.~4
which shows the residual signal to the 2-planet solutions, we can see a
quasi-sinusoidal modulation with the period about of 1000~d. Yet the jitter
estimates of HD~82943 are uncertain by $50\%$ \citep{Lee2006}. Thus by adopting 
values as high as 6--7~m/s, one would obtain $\Chi\sim~1$ and the larger rms 
would be not necessarily unreasonable. Still, the hypothesis about the third 
planet is a very attractive explanation of the rms excess. Below we try to find 
out whether such a configuration would be consistent with a stable dynamics. 

First, we searched for 3-planet  solutions using the
hybrid Kepler code and ``blindly''  assuming the same bounds of the orbital
periods of [10,1200]~d and eccentricities  of [0,0.8] for every planet. The use
of  multi-planet Keplerian model enables us to quickly localize regions of
orbital parameters in which potentially stable, $N$-body fits can be found. The
code was restarted thousands of times. In this search the  algorithm  converged
to a few distinct local minima yielding similar rms and $\Chi$. 

Remarkably, two of the Keplerian best fits, yielding $\Chi \sim 1.2$ and an rms
$\sim 7$~m/s correspond to {\em coplanar} configurations involving two (of
three) Jovian planets in 1:1 MMR. The primary parameters ($K,e,P,\omega,T_0$) of these
fits are given in Table~3 (Fit X and Fit XI, respectively). Their planets c and
d would have similar periods but the eccentricities of the two outer planets  in
1:1 ~MMR are significantly different. Unfortunately, these fits are highly
unstable. We did not succeed in ``stabilizing'' them by  GAMP; nevertheless, the
search was not very extensive and we suspect that stable solutions involving
mutually inclined orbits may exist.

The Fit IX (the mathematically  best fit found in this paper) yields  $\Chi=1.08$
and an rms  $\sim 6.38$~m/s that indicates a almost "perfect" solution.  
It could be  interpreted as a configuration of the
outermost planet accompanying the confirmed  giants involved in the 2:1 MMR.
Unfortunately, this solution is very unstable due to a large eccentricity of the
outermost planet. We tried to refine it with GAMP.  In the relevant range of
semi-major axes  we found stable solutions, and the one we selected is given as
Fit~VIII in Table~2. Let us note that  the stability criterion forces $e_d$ of
this solution to a small value $\sim 0.02$ which   also increases the rms to
about 7.35~m/s. The dynamical maps shown in Fig.~\ref{fig:fig15} reveal  narrow
islands of stability in which the solution is found. The MEGNO signature (the
left-upper panel of Fig.~\ref{fig:fig16}) uncovers a weakly chaotic nature of
this solution. Nevertheless, there is no sign of a physical instability over
at least 
250~Myr (Fig.\ref{fig:fig16} is for the initial 5~Myr integration period). A
peculiarity of this fit is that $e_d$ remains small in spite of a close
proximity to the two larger companions in eccentric orbits.
For a
comparison with the previously found 1:1~MMR and 2:1~MMR solutions, the
synthetic  curve of Fit~VIII is shown
in the bottom panel of Fig.~\ref{fig:fig14}.  

The described results might indicate that our knowledge of \starb system is
still limited in spite of much effort devoted to study the RV data of the parent
star. The currently available  measurements permit many qualitatively different 
orbital solutions well fitting the measurements. Still, the use of stability
criterion  in the fit process seems to be essential to resolve the degeneracy
between very good  but strongly unstable Keplerian 
and Newtonian fits which, as we have shown
above, can easily appear.  

\section{Conclusions}

We have demonstrated the RV measurements for \stara and \starb harboring
2-planet systems may be successfully modeled with two qualitatively different
orbital configuration. One is already recognized configuration corresponding  to
a 2:1 MMR. We show that these observations are equally well modeled with Trojan
pairs of planets (a 1:1 MMR). Both these types of  orbital configurations
produce very similar periodograms of the RV signal.  A common feature of the
Trojan solutions for both systems is the possibility  for large eccentricities
of the orbits, reaching $\sim 0.8$.  Still, the best-fit Trojan configurations
reside in extended zones of rigorously stable quasi-periodic motions. The ease
of maintaining stability and the large zones  of regular motions may strengthen
the hypothesis about the 1:1 MMR configurations. 

It is difficult to explain finding two systems in a 1:1~MMR in the sample of
only  $\sim 20$ multi-planet systems.  A most promising mechanism that might
produce such a configuration is the dynamical relaxation and planetary scattering 
\citep{Adams2003}. In particular, an argument supporting such a hypothesis 
for the HD~82943 system  is the evidence of a planet engulfment by the parent
star \citep{Israelian2001}. That event indicates planetary scaterring in the
past. However, its effect on the currently observed configuration of the system
would be hard to predict. To the best of our knowledge  there are 
no works that could explicitly explain an inclined 1:1~MMR
configuration  as a result of a migration.   On the other hand, the origin of
the 2:1~MMR is a well recognized problem as  we know of at least four systems
presumably involved in such a resonance. Our work demonstrates that the 1:1~MMR
configurations can be used to describe the  observations of \stara and \starb.
This hopefully will encourage others to  study the origin of such
systems.

The best-fit Trojan configurations were found using of our approach  of
modeling the RV data which incorporates a stability indicator. For this purpose
we use a formal KAM criterion which is closely related to  a physical behavior
of a planetary system. This criterion generalizes  the $\max e$ and $\max
\theta$ maps. Still, these maps help us to determine  the character of motions
--- e.g. a type of the corotation  of the apsides. Obviously, all the three
indicators are strictly related.  Presumably, the stability maps would change if
the model of motion included  the relativistic and tidal interactions with the
star.  Although these factors  are orders of magnitude smaller than the leading
gravitational interactions,  their influence might change the overall stability
picture of the systems.  Our work on this subject is ongoing. 

\section{Acknowledgments}
We appreciate discussions with Prof. Sylvio Ferraz-Mello and his help with obtaining
the RV data of the HD~82943 system. We thank Dr. Man Hoi Lee for a detailed
review, critical remarks that improved the manuscript and for providing the full set of
RV measurements of HD~82943. This work is supported by the Polish  Ministry of
Education and Science, Grant No.~1P03D~021~29. M.K. is also supported by NASA
through grant NNG04GM62G.

\appendix

Here we discuss the problem of a proper choice of the stability indicator in a
GAMP-like fitting code. We analyze two 1:1 initial conditions for the \starb 
system which are marked in the stability maps (Fig.~\ref{fig:fig8}, the upper
row; one is our best-fit 1:1 MMR configuration). Let us recall that the 
calculations of $SN$ were conducted over $3\cdot10^4$ orbital periods  of the
planets (about of $6\cdot 10^4$~yr). Apparently, both initial conditions  are
localized in an extended resonance zone. Figure \ref{fig:fig17} (upper row) 
illustrates the temporal MEGNO, $Y(t)$,  as a function of time but computed over
a much longer time-span, $2.5\cdot 10^5$ orbital periods. For the best-fit
solution, the  behavior of $Y(t)$ (oscillations about 2) corresponds to a 
strictly quasi-periodic system while a slow divergence of this indicator can be
observed for the modified initial condition  (marked with a diamond in the $SN$
map, Fig.~\ref{fig:fig8}). It indicates that  in this case the system is in fact
weakly chaotic (the Lyapunov exponent is  relatively small, $\sim
10^{-5}~\mbox{yr}^{-1}$). An inspection of  the $\max \theta$-map for this
initial condition reveals that $\theta$ in this  case circulates
(Fig.~\ref{fig:fig8} and Fig.~\ref{fig:fig17}, right column)  but on the average
the apsides are anti-aligned and this helps to maintain the stability. It means
that $\max \theta$ would not be a good choice as a stability indicator.
Nevertheless, it is a valuable tool for resolving the complex structure of the
resonance.

%
%

%
%
%
{
\begin{table}
\caption{\rm
The best-fit 2-planet initial conditions for the \stara  \citep{Vogt2005} and
\starb \cite{Mayor2004} planetary systems found with GAMP (MEGNO was calculated
over $\simeq 1000$--$5000$ periods of the more distant companion). Jitter
estimates are $9$~m/s for \stara and $5$~m/s for \starb. Astrocentric osculating
elements  are given for the date of the first observation from \cite{Vogt2005}
and \cite{Mayor2004}, respectively. 
The masses of the parent stars are of
$0.84\mbox{M}_{\sun}$ for \stara and of $1.15~\mbox{M}_{\sun}$ for \starb.
}
\smallskip
\centering
\begin{tabular}{lcccccccc}
\hline
  			& 	\multicolumn{2}{c}{\ Fit I\ } &
				\multicolumn{2}{c}{\ Fit II\ } &
				\multicolumn{2}{c}{\ Fit III\ } &
				\multicolumn{2}{c}{\ Fit IV \ } \\
  			& 	\multicolumn{2}{c}{\ HD 128311 (2:1 MMR)\ } &
				\multicolumn{2}{c}{\ HD 128311 (1:1 MMR)\ } &
				\multicolumn{2}{c}{\  HD 82943 (2:1 MMR)\ } &
				\multicolumn{2}{c}{\  HD 82943 (1:1 MMR)\ } \\
Orbital parameter \hspace{1em}  & \ \ {\bf b}  \ \ & \ \ {\bf c} \ \ 
			        & \ \ {\bf b}  \ \ & \ \ {\bf c} \ \ 
			        & \ \ {\bf b}  \ \ & \ \ {\bf c} \ \ 
				& \ \ {\bf b} \ \ & \ \  {\bf c} \ \ 
\\
\hline
$\mbox{m}_2 \sin i$ [M$_{\idm{J}}$] \dotfill 
				&  1.606 &  3.178  
				&  7.174 &  6.954  
				&  1.810 &  1.812 
				&  9.888 &  4.182      
\\
a [AU] \dotfill 		&  1.112 &  1.732
				&  1.737 &  1.796 
				&  0.744 &  1.192
				&  1.187 &  1.201
\\
e \dotfill     			&  0.359 &  0.214
				&  0.311 &  0.599
				&  0.395 &  0.128
				&  0.504 &  0.658
\\				
$i$ [deg] \dotfill		& 90.00  &  90.00
				& 44.22  &  16.96 
				&  90.00 &   90.00
				& 19.23  &  19.57			       
\\
$\omega$ [deg]\dotfill 		&  71.58 &  12.71
				&  84.14 &  112.53 
				&  121.25&  222.94
				&  123.74&  126.33
\\
$\Omega$ [deg]\dotfill 		&  0.0   &  0.0
				& 209.79 &  0.0   
				&  0.0   &  0.0 
				& 178.52 &  0.0
\\
$M(t_0)$ [deg] \dotfill 	& 271.72 &  190.23
				& 125.24 &  311.56
				& 355.99 &  258.96				
				& 356.02 &  170.30
\\
$V_0$ [m/s] \dotfill 	& 	\multicolumn{2}{c}{0.970} &
				\multicolumn{2}{c}{0.655} &
				\multicolumn{2}{c}{-0.761} &
				\multicolumn{2}{c}{-2.877} 
\\ 
$\Chi$  \dotfill 	&       \multicolumn{2}{c}{1.731} & 
		 		\multicolumn{2}{c}{1.797} &
		 		\multicolumn{2}{c}{1.047} &
		 		\multicolumn{2}{c}{1.221} 
\\
rms [m/s] \dotfill 	& 	\multicolumn{2}{c}{ 15.28}  & 
				\multicolumn{2}{c}{ 15.49} &
				\multicolumn{2}{c}{ 7.12} &
				\multicolumn{2}{c}{ 8.13}
\\
\hline
\end{tabular}
\label{tab:tab1}
\end{table}
}

%
%
%
{
\begin{table}
\caption{\rm
The best-fit 2-planet initial conditions for the \starb planetary system, on the
basis of  data set used by \cite{Lee2006}. The stable fits are found  with GAMP
(MEGNO was calculated over $\simeq 1000$--$5000$ periods of the more distant
companion). Jitter estimate is $4.2$~m/s. Astrocentric osculating elements  are
given for the date of the first observation from  \cite{Mayor2004}. The mass of
the parent star is  $1.15~\mbox{M}_{\sun}$. CORALIE RV data are shifted by
8128.598~m/s.
}
\smallskip
\centering
\begin{tabular}{lccccccccc}
\hline
  			& 	\multicolumn{2}{c}{\ Fit V (stable)\ } &
				\multicolumn{2}{c}{\ Fit VI (stable)\ } &
				\multicolumn{2}{c}{\ Fit VII (unstable)\ } &
				\multicolumn{3}{c}{\ Fit VIII (stable)\ } 
\\
  			& 	\multicolumn{2}{c}{\ HD 82943 (2:1 MMR)\ } &
				\multicolumn{2}{c}{\ HD 82943 (1:1 MMR)\ } &
				\multicolumn{2}{c}{\ HD 82943 (2:1 MMR)\ } &
				\multicolumn{3}{c}{\ HD 82943 (3-planet)\ } 
\\
Orbital parameter \hspace{1em}  & \ \ {\bf b}  \ \ & \ \ {\bf c} \ \ 
			        & \ \ {\bf b}  \ \ & \ \ {\bf c} \ \ 
			        & \ \ {\bf b}  \ \ & \ \ {\bf c} \ \ 
			        & \ \ {\bf b}  \ \ & \ \ {\bf c} \ \ 
				& \ \ {\bf d}   \ \ 
\\
\hline
$\mbox{m}_2 \sin i$ [M$_{\idm{J}}$] \dotfill 
				&  1.461 &  1.728
				&  2.043 &  3.932
				&  17.16 &  1.761
				&  1.679 &  1.867 & 0.487    
\\
a [AU] \dotfill 		&  0.748 &  1.186
				&  1.208 &  1.180
				&  0.751 &  1.240
				&  0.751 &  1.197 & 2.125
\\
e \dotfill     			&  0.448 &  0.268
				&  0.640 &  0.500
				&  0.380 &  0.001
				&  0.386 &  0.110 & 0.018
\\				
$i$ [deg] \dotfill		&  90.00  & 90.00 
				&  49.35  & 56.56 
				&  6.260  & 87.85
				&  90.0 &  90.0 & 90.0		       
\\
$\omega$ [deg]\dotfill 		& 126.82 &  138.35
				& 133.92 &   127.88
				& 119.84 &   187.81
				& 118.08 &  144.47 & 114.61
\\
$\Omega$ [deg]\dotfill 		& 0.0    &  0.0
   		                & 145.71  &  0.0
				& 346.23 &  0.0   
				& 0.0   &  0.0  & 0.0
\\
$M(t_0)$ [deg] \dotfill 	& 359.23 &  336.85
				& 186.39       &  353.84
				& 0.00       &  286.03
				&  2.65 &  345.24 & 79.76				
\\
$V_0$ [m/s] \dotfill 	& 	\multicolumn{2}{c}{13.66} &
				\multicolumn{2}{c}{12.41} &
				\multicolumn{2}{c}{15.42} &
				\multicolumn{3}{c}{14.60}
\\
$V_1$ [m/s] \dotfill 	& 	\multicolumn{2}{c}{-7.72} &
				\multicolumn{2}{c}{-6.86} &
				\multicolumn{2}{c}{-4.96} &
				\multicolumn{3}{c}{-0.73}
\\				
$\Chi$  \dotfill 	&       \multicolumn{2}{c}{1.39} & 
		 		\multicolumn{2}{c}{1.45} &
		 		\multicolumn{2}{c}{1.32} &
		 		\multicolumn{3}{c}{1.27} 
			
\\
rms [m/s] \dotfill 	& 	\multicolumn{2}{c}{7.98} & 
				\multicolumn{2}{c}{8.40} &
				\multicolumn{2}{c}{7.58} &
				\multicolumn{3}{c}{7.36} 
\\
\hline
\end{tabular}
\label{tab:tab2}
\end{table}
}

{
\begin{table}
\label{tab:tab3}
\caption{\rm
Primary best-fit parameters of the 3-planet Kepler models found in this paper on
the basis of RV measurements of HD~82943 used by \cite{Lee2006}. The epoch $T_0$
is JD~2,450,000. 
The adopted jitter estimate is 4.2~m/s.  CORALIE RV measurements  are
shifted by 8128.598~m/s. All fits are dynamically unstable.
}
\centering
\begin{tabular}{lcccccccccc}
\hline
&\multicolumn{3}{c}{Fit IX} 
&\multicolumn{3}{c}{Fit X}
&\multicolumn{3}{c}{Fit XI} 
\\
  			& 	\multicolumn{3}{c}{\ HD 82943 (3-planet)\ } &
				\multicolumn{3}{c}{\ HD 82943 (3-planet)\ } &
				\multicolumn{3}{c}{\ HD 82943 (3-planet)\ } 
\\
Parameter \hspace{1em}  
& \ \  {\bf b} \ \  & \ \ {\bf c} \ \ & \ \  {\bf d} \ \  
& \ \  {\bf b} \ \  & \ \ {\bf c} \ \ & \ \  {\bf d} \ \  
& \ \  {\bf b} \ \  & \ \ {\bf c} \ \ & \ \  {\bf d} \ \  
\\
\hline
$K$ [m/s] \dotfill 		&   59.735   &  41.838  & 10.493
			&   55.926   &  16.997  & 36.487
                        &   51.173   &  19.853  & 36.059
\\
$P$ [d] \dotfill 		&  219.423  &  442.893  & 937.663
                        &  219.766  &  417.579  & 445.914
                        &  219.536  &  418.197  & 449.093
\\
$e$  \dotfill   		&  0.398   &   0.141  &  0.580
			&  0.403   &   0.712  &  0.061
                        &  0.437   &   0.683  &  0.210
\\
$\omega$ [deg] 	\dotfill	& 107.386    &  86.565  & 215.039
			& 120.701    &  240.476  & 100.133
                        & 119.385    &  236.421 & 97.881
\\
$T_{\idm{p}}$ [JD-$T_0$] \dotfill
			& 1842.338  &  232.810  & 384.940
			& 2505.152  &  2141.264   &  3819.642
                        & 3163.810  &  1722.905 & 1585.272
\\
$\Chi$  \dotfill		& \multicolumn{3}{c}{1.079} & 
		 	  \multicolumn{3}{c}{1.183} &
		 	  \multicolumn{3}{c}{1.180} 
\\
rms~[m/s] \dotfill		& \multicolumn{3}{c}{6.37}  & 
			  \multicolumn{3}{c}{6.97}  & 
			  \multicolumn{3}{c}{7.01} 
\\
$V_0$ [m/s] \dotfill		& \multicolumn{3}{c}{-3.80}  &
			  \multicolumn{3}{c}{-8.90} &
			  \multicolumn{3}{c}{-8.65} 
\\
$V_1$ [m/s] \dotfill		& \multicolumn{3}{c}{17.06}  &
			  \multicolumn{3}{c}{16.99} &
			  \multicolumn{3}{c}{16.77} 
\\ 
\hline
\end{tabular}
\end{table}
}

\clearpage

%
%
%

%
%

\figcaption[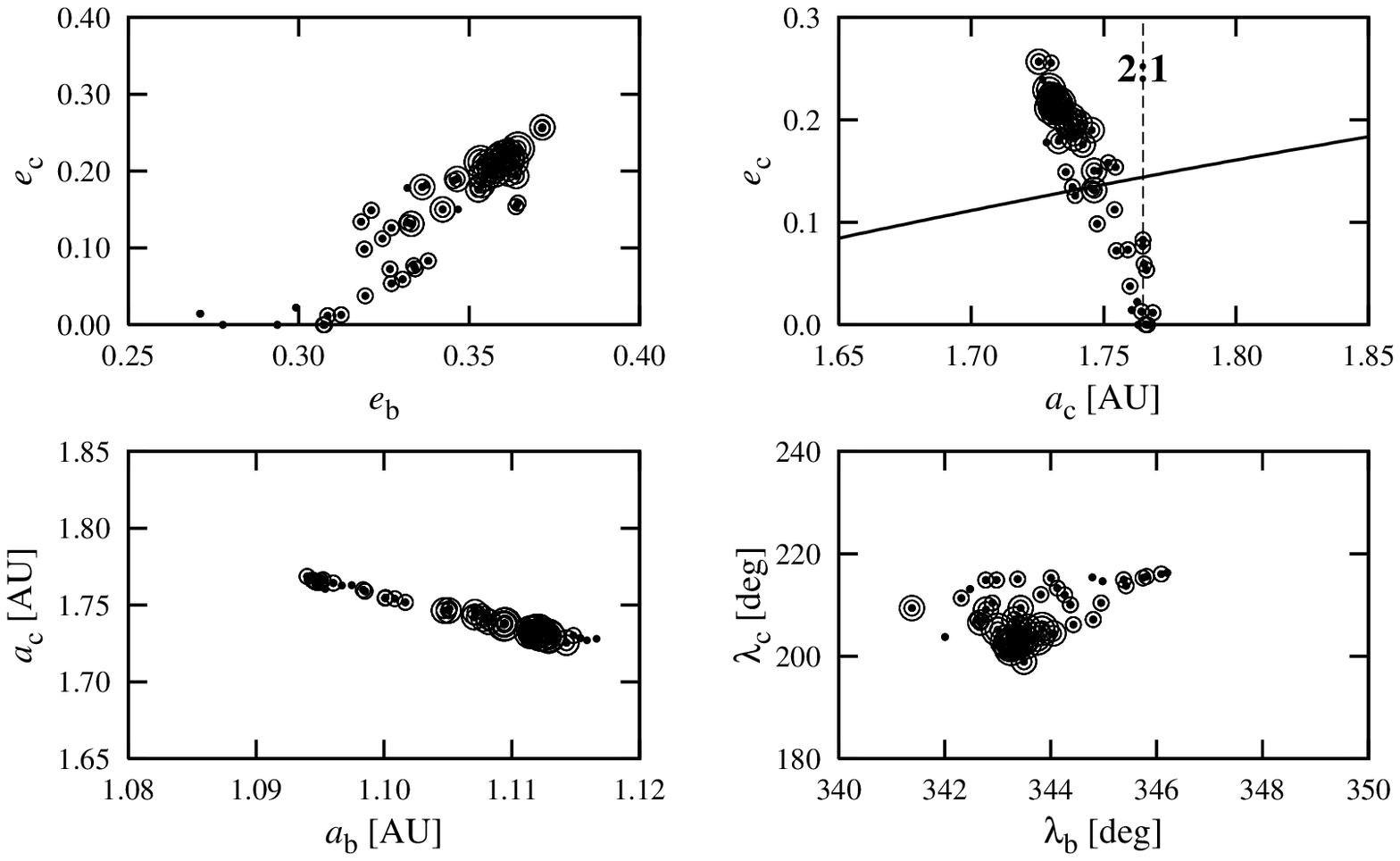]{\normalsize
The best fits obtained by the GAMP algorithm for the RV data published in
\cite{Vogt2005} for HD~128311. In the model, the coplanar system and 2:1 MMR is
assumed. Parameters of the fit are projected onto the planes of osculating
orbital elements. The smallest filled circles are for the solutions  with $\Chi$
within the formal $3\sigma$ confidence interval of the best-fit solution, with
$\Chi< 1.79$. Bigger open circles are for $\Chi<1.761$  and $\Chi< 1.741$
($2\sigma$ and $1\sigma$ confidence intervals of the best-fit solution,
respectively). The largest circles are for the solutions with $\Chi< 1.732$
marginally larger than $\Chi=1.731$ of the best-fit  initial condition (Fit I,
Table 1). A curve  in the $(a_{\idm{c}},e_{\idm{c}})$-plane denotes the
planetary collision line which  is determined from the relation 
$a_{\idm{b}}(1+e_{\idm{b}})=a_{\idm{c}}(1-e_{\idm{c}})$  with
$a_{\idm{b}},e_{\idm{b}}$  fixed at their best-fit values. The nominal  position
of the  2:1 MMR inferred from the Kepler law is also marked.
}

%
%
%

\figcaption[f2a.eps,f2b.eps,f2c.eps,f2d.eps]{\normalsize
The stability maps in the $(a_{\idm{c}},e_{\idm{c}})$-plane in terms of the
Spectral Number, $\log SN$, $\max e$ and $\max \theta$, for the best-fit 
solution corresponding to the putative 2:1 MMR for the coplanar HD~128311 
system (see Table~1, Fit I). Colors used in the $\log SN$ map classify the
orbits --- black indicates quasi-periodic, regular configurations while yellow
strongly chaotic systems. A circle denotes the best-fit configuration
related to Fit~I. The
resolution of the maps is $600\times120$ data points. Integrations are for
$3\cdot 10^4$ periods of the outer planet ($\sim 7\cdot 10^4)$~yr.
}

%
%
%

\figcaption[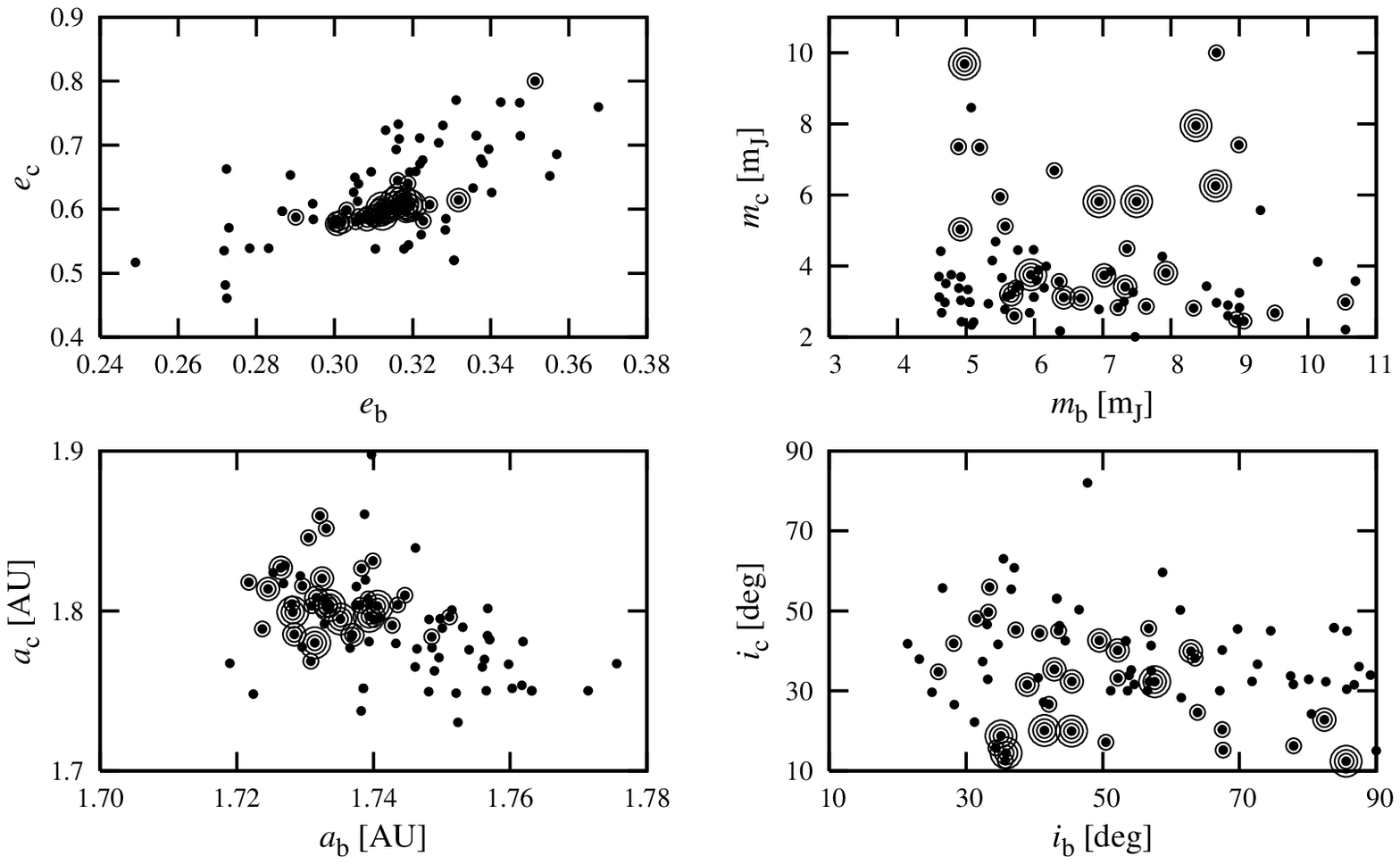]{\normalsize
The solutions obtained with GAMP for the RV data from \cite{Vogt2005} for
HD~128311. In the model, an inclined system and a 1:1 MMR is assumed. Orbital
parameters are projected onto the planes of osculating elements. The smallest
filled circles are for solutions with $\Chi$ within the formal $3\sigma$
confidence interval of the best-fit (Table~1; $\Chi<1.9$ and the rms about
17~m/s). Bigger open circles are for  $\Chi<1.825$ and $\Chi<1.81$ ($2\sigma$
and $1\sigma$ confidence intervals, respectively). The largest circles are for
the solutions with $\Chi< 1.799$ marginally larger from $\Chi=1.797$ of the
best-fit solution given in Table~1, Fit II. Compare the formal $3\sigma$ range
for the solutions shown in the $(a_{\idm{b}},a_{\idm{c}})$-plane  with the width
of the 1:1~MMR for the best-fit solution (Fig.~\ref{fig:fig6}, bottom row).
}

%
%
%

\figcaption[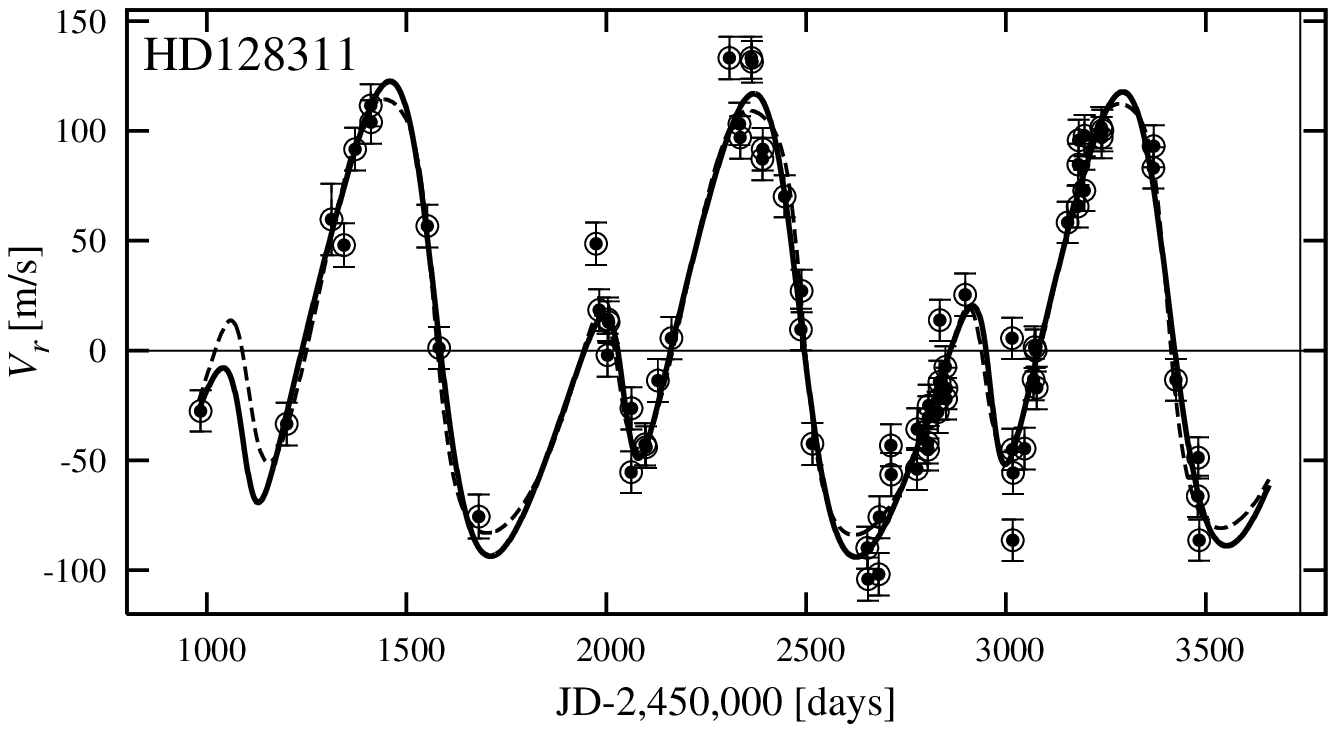]{\normalsize
The synthetic RV curves for the best-fit solutions corresponding to  the 2:1 and
1:1 MMRs in the HD~128311 planetary system (see also Table~1, Fits I and II). 
Thick line is for the 1:1 MMR and dashed line is for the 2:1 MMR. Both curves 
give an rms of $\simeq 15$~m/s. The error bars include the stellar jitter  of
9~m/s.   
}

%
%
%

\figcaption[f3.eps]{\normalsize
The Lomb-Scargle periodogram for the best fit solutions 
found for the \stara{} system (Fit I and II in Table 1). 
The thick line is  for 
the synthetic RV corresponding to the 2:1~MMR. The thin line is for the RV curve 
of the 1:1~MMR solution. The dashed line is for the measurements.}

%
%
%

\figcaption[f6a.eps,f6b.eps,f6c.eps,f6d.eps,f6e.eps,f6f.eps]{\normalsize
The stability maps in the $(e_{\idm{b}},e_{\idm{c}})$ (upper row, the
resolution  is $250\times250$ data points) and  $(a_{\idm{c}},e_{\idm{c}})$
plane (lower row,  the resolution is $480\times200$ data points)
for \stara (Fit II). The left
column is for the Spectral  Number, $\log SN$.  Colors used in the $\log SN$ map
classify the orbits --- black indicates quasi-periodic, regular configurations
while yellow strongly  chaotic systems. The maps marked with $\max e_{\idm{c}}$
and $\max {\theta}$ are respectively  for the maximal eccentricity and the
maximum of $\theta=\varpi_{\idm{b}}-\varpi_{\idm{c}}$ attained during the
integration of the system. A  circle marks the parameters of the best-fit
solution. The integration was conducted for $\sim 6\cdot 10^4$ orbital periods
of the planets.
}

%
%
\figcaption[f7a.eps,f7b.eps,f7c.eps]{\normalsize
The stability maps in the $(e_{\idm{b}},e_{\idm{c}}$)-plane (the resolution is
$250\times250$ data points) for the fit of the 1:1 MMR in the \stara system
having a slightly
larger  rms than the best-fit solution (Fit II in Table~1), $\simeq 15.7$
m/s and $\Chi=1.82$. The osculating  elements at the date of the first
observation are   $(m~\mbox{[m$_\idm{{J}}$]},a~\mbox{[AU]},e,i~\mbox{[deg]},
\Omega~\mbox{[deg]},\omega~\mbox{[deg]},M~\mbox{[deg]})$:
(7.22, 1.730, 0.323, 43.52, 220.38, 80.18, 129.99) for the planet~b and
(2.83, 1.816, 0.582, 45.00, 0.00, 113.43, 312.87) for the planet~c; $V_0=-0.078$~m/s.
The left column is for the spectral number, $\log SN$. Colors used in the $\log
SN$ map classify the orbits --- black indicates quasi-periodic, regular
configurations while yellow strongly chaotic systems. The maps marked by $\max
e_{\idm{c}}$ and $\max {\theta}$ are respectively for the maximal eccentricity
of the outermost planet and the maximum of $\theta=\varpi_{\idm{b}}-\varpi_{\idm{c}}$  attained during
the integration of the system. A circle marks the parameters of the best-fit
solution. The integration was conducted for $\sim 6\cdot 10^4$ orbital periods
of the planets. 
}

%
%
%

\figcaption[f8a.eps,f8b.eps,f8c.eps,f8d.eps,f8e.eps,f8f.eps]{\normalsize
The stability maps in the $(e_{\idm{b}},e_{\idm{c}})$ (upper row, the resolution
is $250\times250$ data points) and $(a_{\idm{c}},e_{\idm{c}})$ plane (lower row,
the resolution is $400\times200$ data points) for \starb (Fit IV, see Table~1).
The left  column is for the spectral number, $\log SN$. Colors used in the $\log
SN$  map classify the orbits --- black indicates quasi-periodic, regular 
configurations while yellow strongly chaotic systems. The maps marked by $\max
e_{\idm{c}}$ and $\max {\theta}$ are respectively  for the maximal eccentricity
and the maximum of $\theta=\varpi_{\idm{b}}-\varpi_{\idm{c}}$ attained during
the integration of the system. A  circle marks the parameters of  the best-fit
solution corresponding to the 1:1 MMR in the \starb system  (Fit IV, Table~1). 
The integration was conducted for $\sim 6\cdot 10^4$ orbital periods of the
planets. The diamond at ($e_{\idm{b}}=0.4,e_{\idm{c}}=0.2$)
is for the initial condition in  a discussion of the
proper choice of the stability indicator in the  GAMP-like code (see the
Appendix).
}

%
%
%
\figcaption[f9a.eps,f9b.eps,f9c.eps,f9d.eps,f9e.eps,f9d.eps]{\normalsize
The stability maps in the $(e_{\idm{c}},\omega_{\idm{c}})$-plane of the \starb{}
system  (the resolution is $240\times240$ data points) for the 2-planet edge-on
best fit solutions related to the 2:1~MMR in 
the HD~82943 system. The top row is for the best stable Fit V
(Table~2). 
The bottom row is for the alternative marginally worse solution
with the following astrocentric elements ($m_p\sin i,a,e,\omega,M$)
at the epoch of the first observation: (1.781~m$_{\idm{J}}$, 0.749~AU,  0.399,
118.273$^\circ$, 0.000$^\circ$) and (1.773~m$_{\idm{J}}$, 1.194~AU,  0.012,
261.882$^\circ$, 221.483$^\circ$) for the inner and outer planet, respectively;
an rms of this fit is $\sim 8.1$~m/s.
The middle row is for 2-planet Keplerian Fit II of \cite{Lee2006}.
The left column is for the Spectral
Number, $\log SN$. The colors used in the $\log SN$ map classify the orbits ---
black indicates quasi-periodic regular configurations while yellow strongly
chaotic ones. The maps in the right column and marked with $\max e_{\idm{c}}$ are for 
the maximal eccentricity of the outermost planet attained during the integration of the system. 
The circle marks the parameters of the best-fit solutions. 
The integrations were
conducted for $\sim 4\cdot 10^4$ orbital periods of the outermost planet.
}

%
%
%
\figcaption[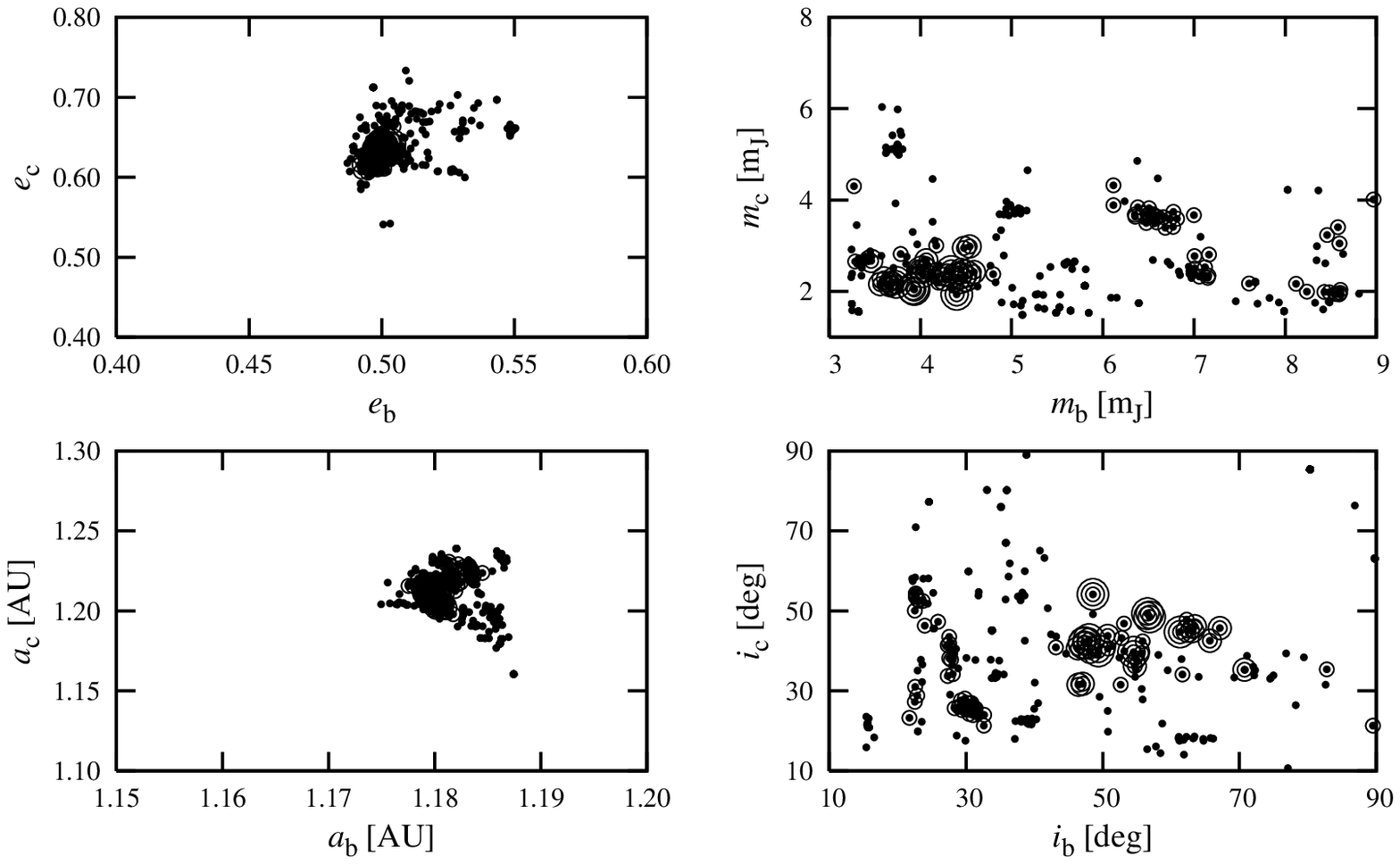]{\normalsize
The solutions obtained with GAMP for the RV data from \cite{Lee2006} for
HD~82943. In the model, mutually inclined orbits and  the presence of the 1:1
MMR is assumed. Orbital parameters are projected onto the planes of osculating
elements. The smallest filled circles are for solutions with $\Chi$ within the
formal $3\sigma$ confidence interval of the best-fit (Fit VI in Table~2);
$\Chi<1.55$ and the rms about 9~m/s. Bigger open circles are for  $\Chi<1.46$
and $\Chi<1.45$ ($2\sigma$ and $1\sigma$ confidence intervals, respectively).
The largest circles are for the solutions with $\Chi< 1.449$ marginally larger
from $\Chi=1.447$ of the best-fit VI given in Table~2. 
}

%
%
\figcaption[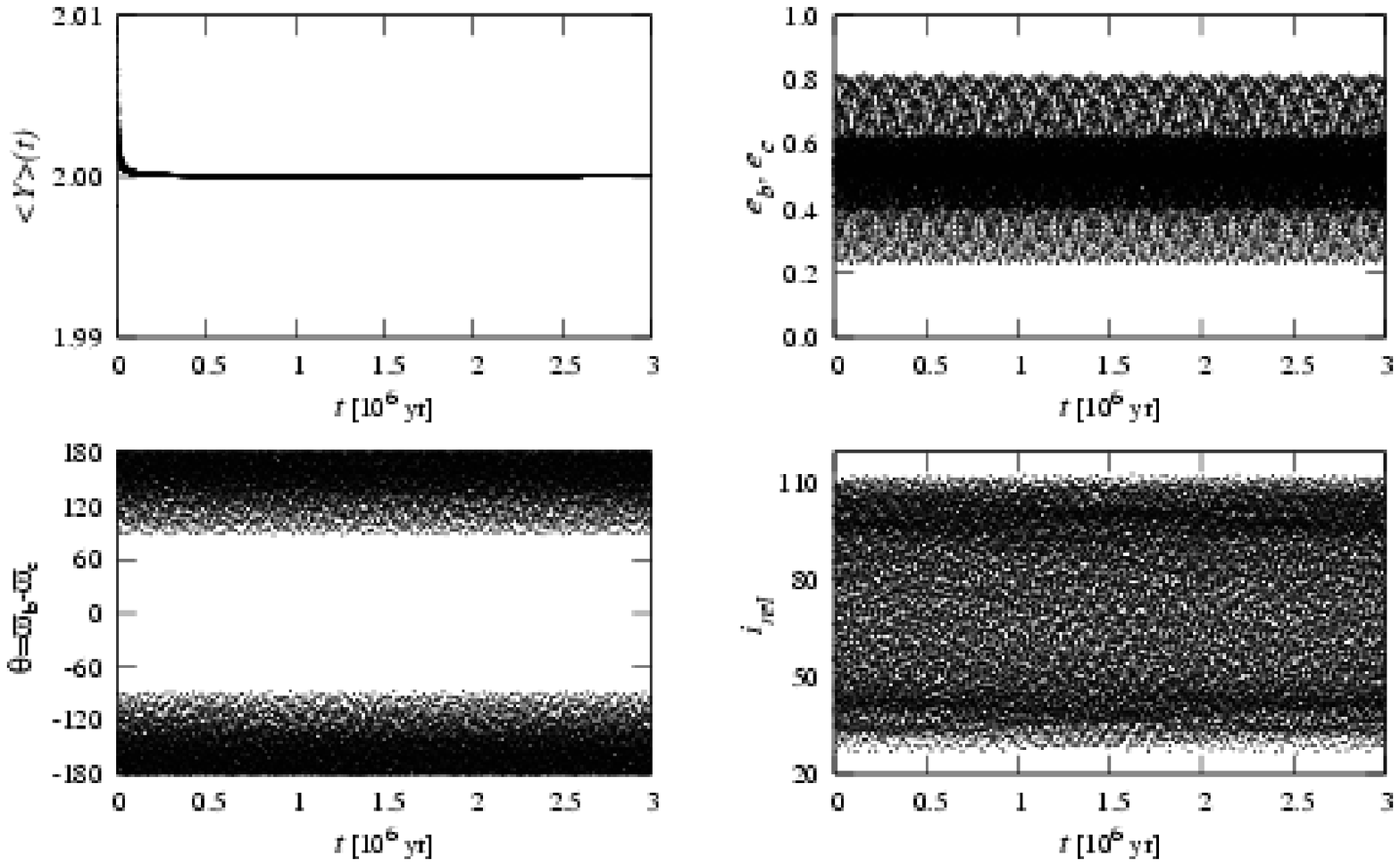]{\normalsize
Evolution of MEGNO and orbital elements of the configuration 
described by Fit~VI in Table~2. A perfect convergence of MEGNO over $\sim 3$~Myr
indicates a rigorously stable solution. Subsequent panels are for the eccentricities,
the critical argument of secular resonance $\theta$ and the relative inclination
of orbits, $i_{\idm{rel}}$.
}

%
%
%
\figcaption[f12a.eps,f12b.eps,f12c.eps,f12d.eps,f12e.eps,f12f.eps]{\normalsize
The stability maps in the $(e_{\idm{b}},e_{\idm{c}})$ (upper row, the resolution
is $250\times240$ data points) and $(a_{\idm{c}},e_{\idm{c}})$ plane (lower row,
the resolution is $300\times100$ data points) for \starb (Fit VI, see Table~2).
The left  column is for the spectral number, $\log SN$. Colors used in the $\log
SN$  map classify the orbits --- black indicates quasi-periodic, regular 
configurations while yellow strongly chaotic systems. The maps marked by $\max
e_{\idm{c}}$ and $\max {\theta}$ are respectively  for the maximal eccentricity
and the maximum of $\theta=\varpi_{\idm{b}}-\varpi_{\idm{c}}$ attained during
the integration of the system. A  circle marks the parameters of  the best-fit
solution corresponding to the 1:1 MMR in the \starb system  (Fit VI, Table~2). 
The integration was conducted for $\sim 3\cdot 10^4$ orbital periods of the
planets. 
}

%
%
%
\figcaption[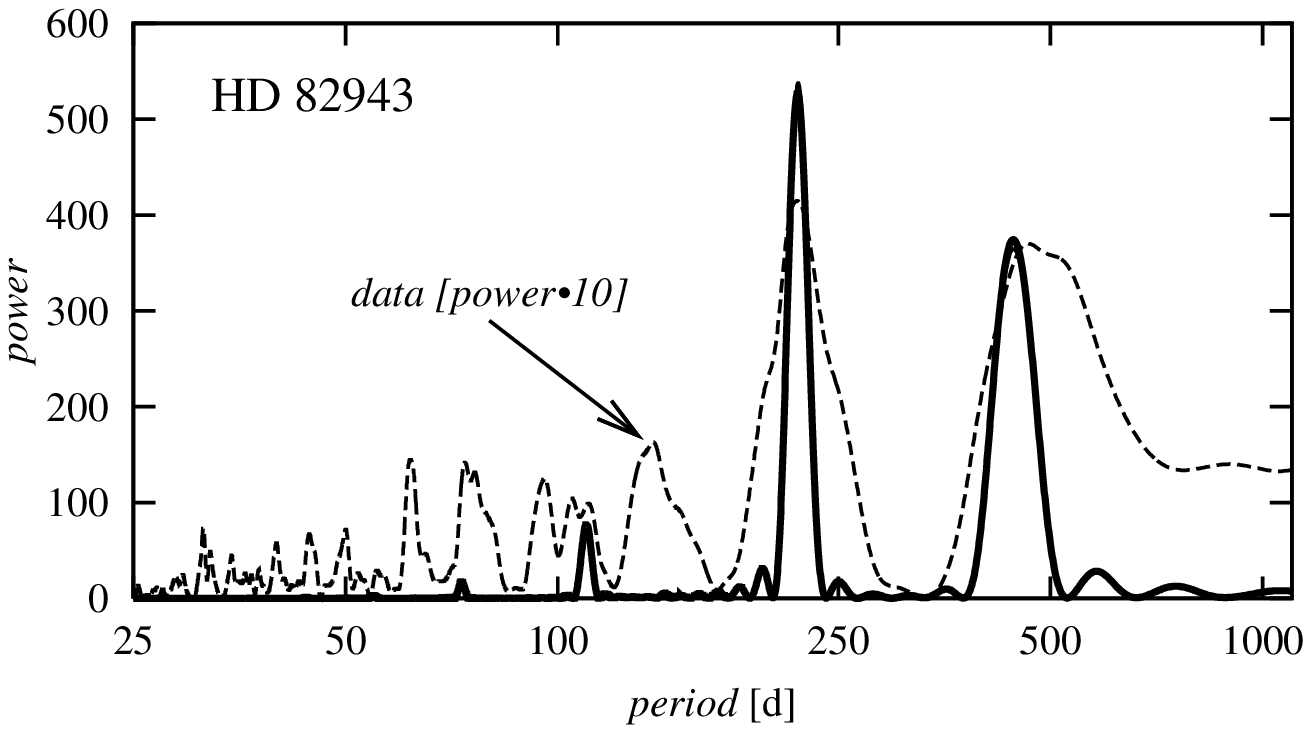]{\normalsize
The Lomb-Scargle periodogram for the best fit solutions (Fit V and Fit VI, Table
2) found for the \starb{} system. The thick line is for the synthetic RV
corresponding to the 2:1~MMR. The thin line is for the RV curve of the 1:1~MMR
solution. The dashed line is for the measurements from \cite{Lee2006}.
}

%
%
%
\figcaption[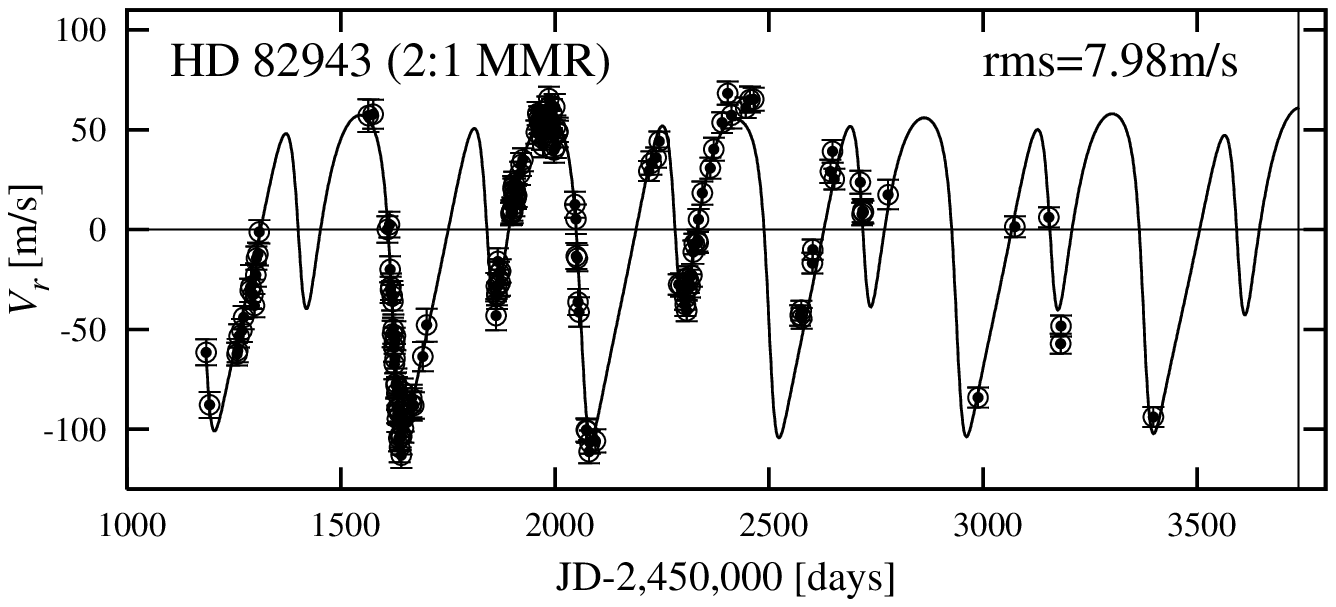,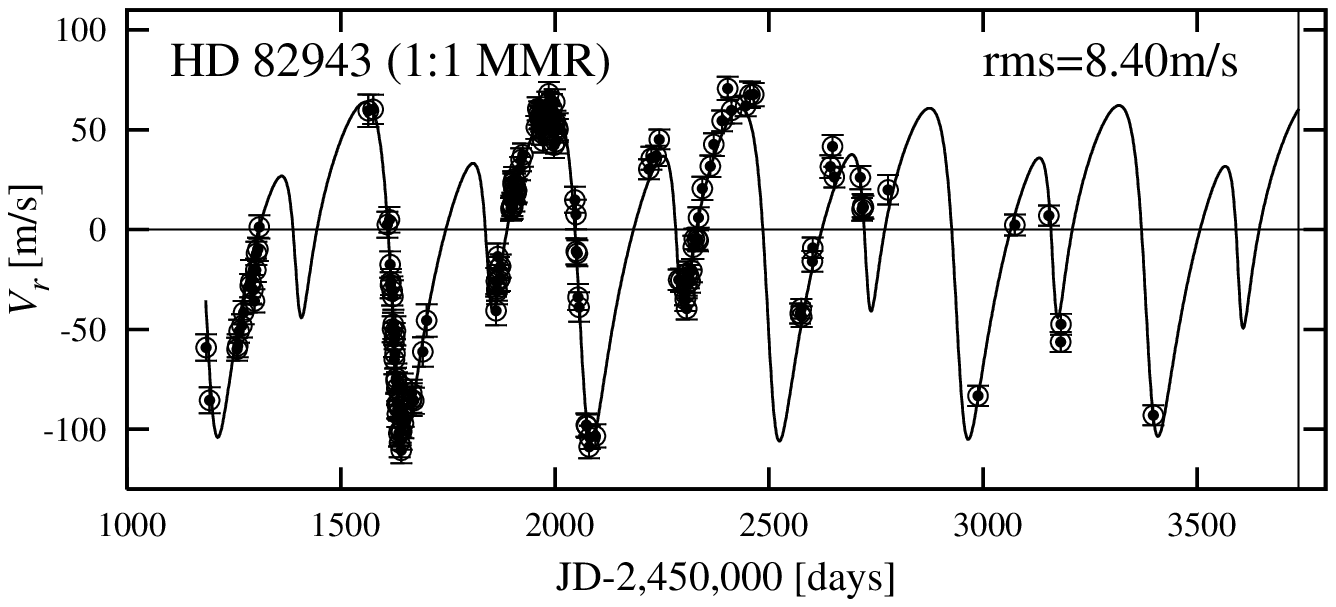,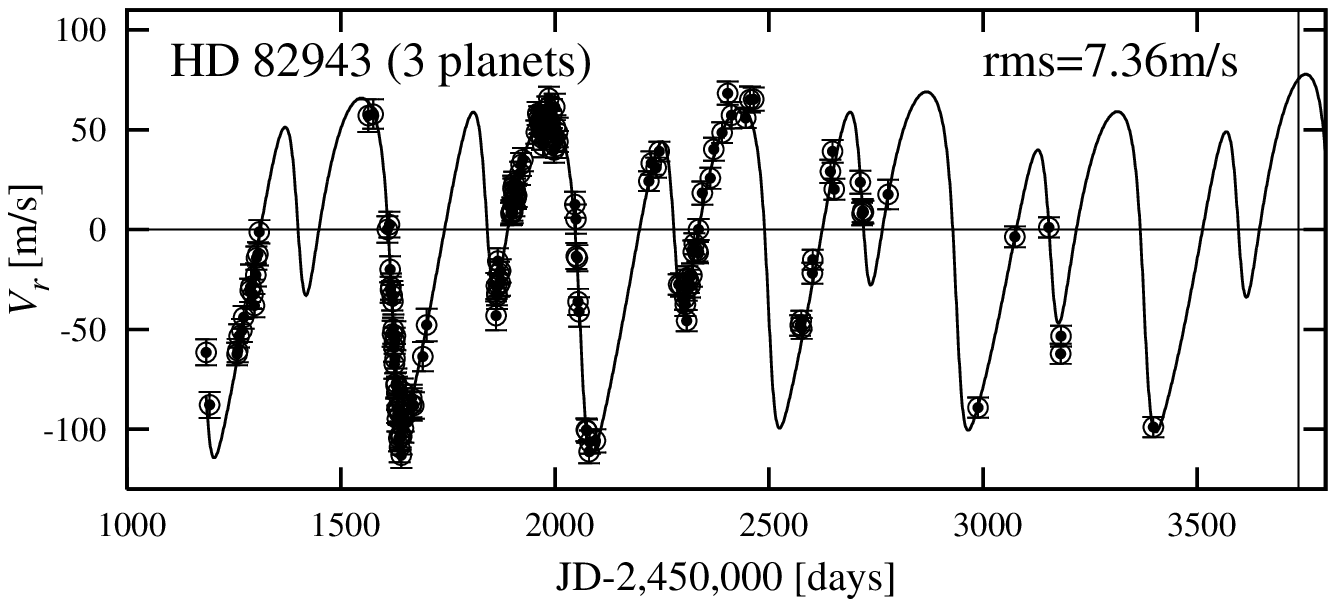]{\normalsize
The synthetic RV curves for the \starb{} system. 
The top plot is for a stable
($N$-body) solution corresponding to a 2:1~MMR (Fit~V).
The middle plot is for
the 1:1~MMR solution (Fit~VI).
The bottom plot
is for stable Newtonian, 3-planet  best fit solution
(Fit~VIII).  The open circles are for the RV measurements from \cite{Lee2006}.
The error bars include stellar jitter of 4.2~m/s.
}

%
%
%
\figcaption[f15c.eps,f15b.eps]{\normalsize
The stability maps in the $(a_{\idm{d}},e_{\idm{d}})$-plane of the \starb{}
system  (the resolution is $300\times120$ data points) for the 3-planet
Newtonian Fit VIII (Table~2). 
The left panel is for the Spectral  Number, $\log SN$. Colors
used in the $\log SN$ map classify the orbits --- black indicates
quasi-periodic regular configurations while yellow strongly chaotic ones. 
The right panel marked with $\max e_{\idm{d}}$ is for  the maximal
eccentricity  attained during the integration of the system. The circle marks
the parameters of the best-fit solution (Fit VIII, Table~2). 
The integration was
conducted for $\sim 6\cdot 10^4$ orbital periods of the outermost planet.
}

%
%
\figcaption[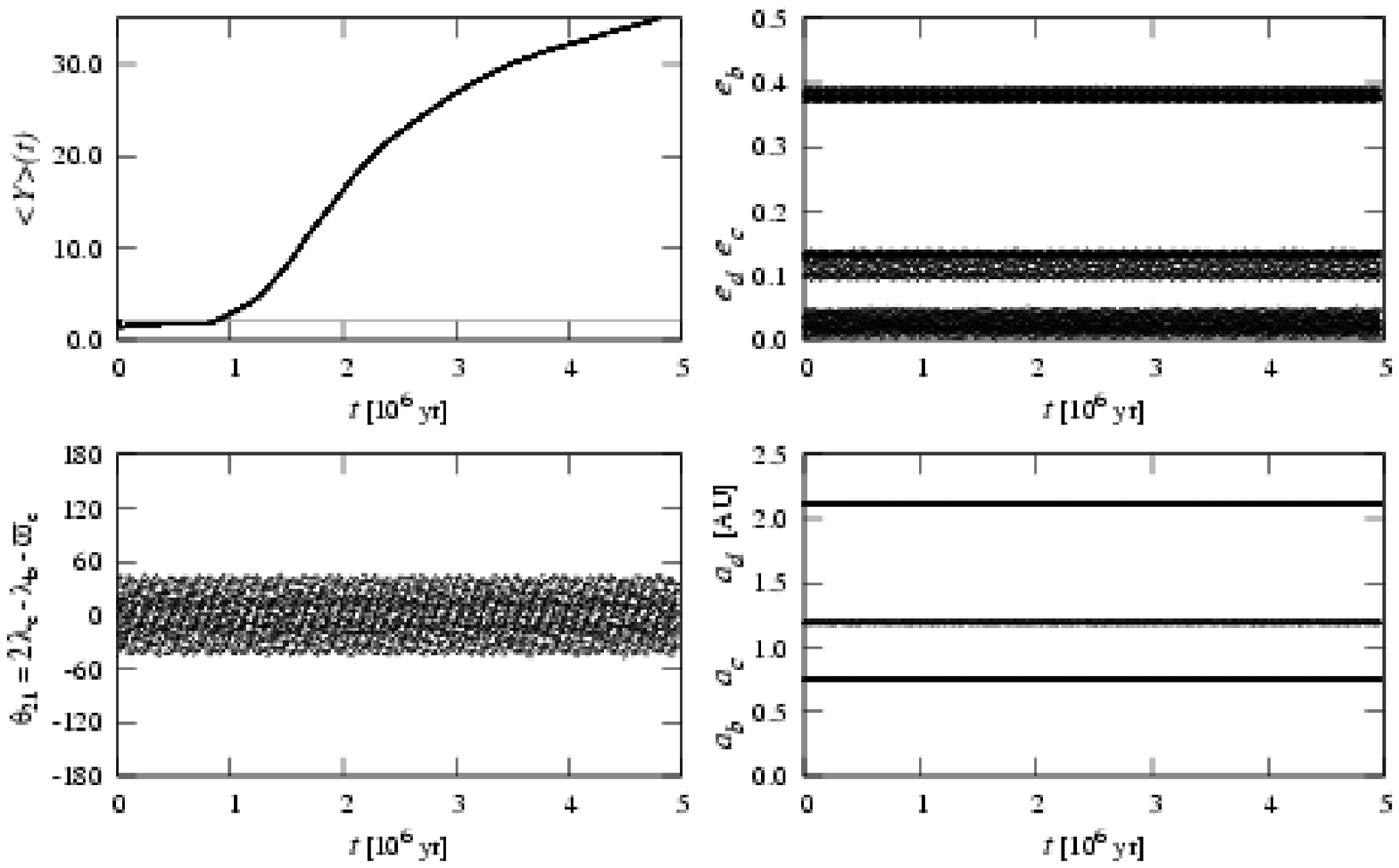]{\normalsize
Evolution of MEGNO and orbital elements of the 3-planet  configuration
described by Fit~VIII in Table~2. A slow divergence of MEGNO after $\sim 1$~Myr
indicates a marginally unstable solution. The evolution of the elements does not
change over at least 250~Myr (not shown here). The subsequent panels are for the
eccentricities, the critical angle of the 2:1 MMR and the semi-major axes.
}

%
%
\figcaption[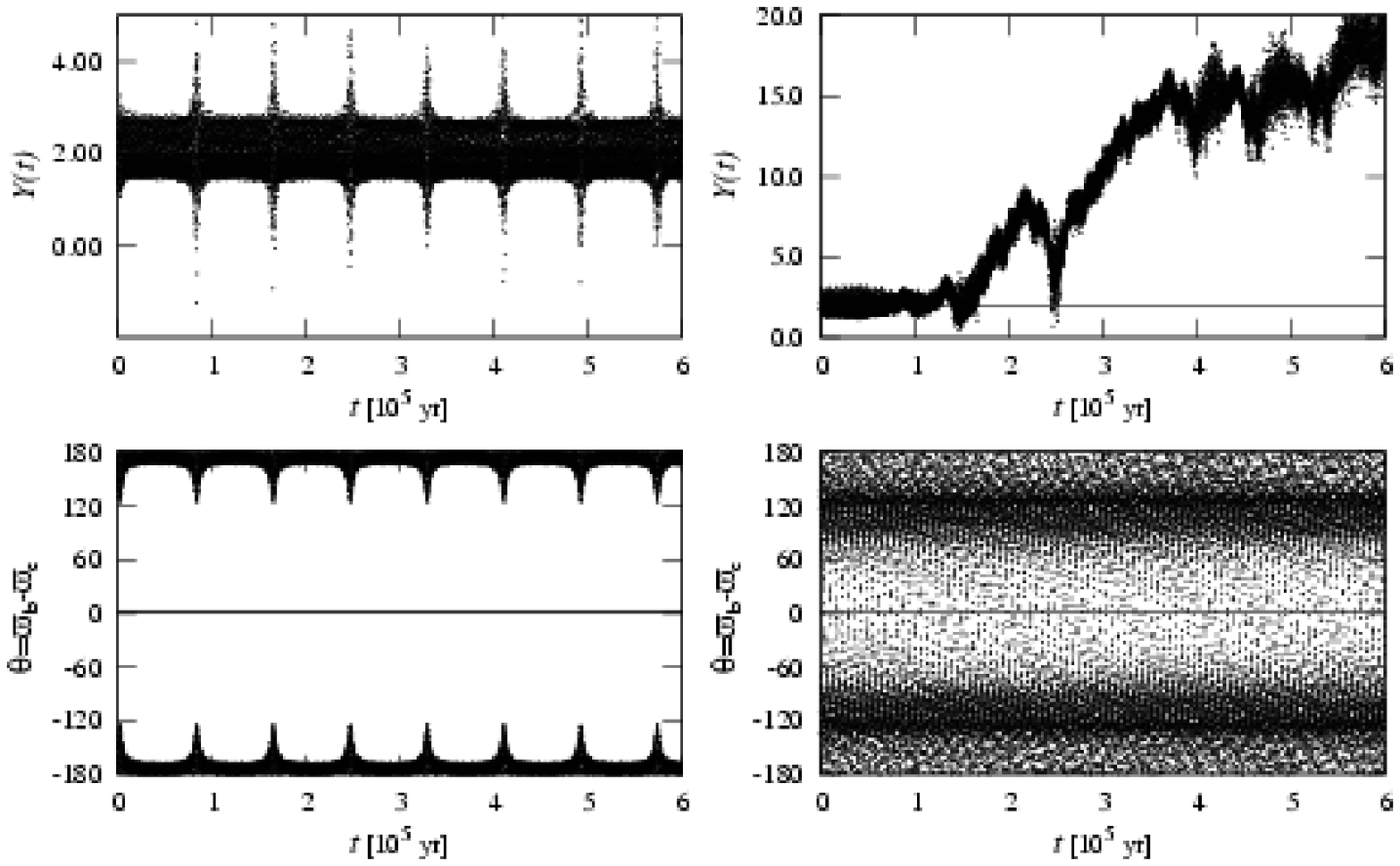]{\normalsize
The left column is for the evolution of MEGNO, $Y(t)$, and $\theta$ for  the
best-fit 1:1 MMR solution (Fit~IV, Table~1) for \starb. The right column is for
the  initial condition marked with a diamond in Fig.~\ref{fig:fig8}.
}

\setcounter{figure}{0}

%
%
\begin{figure*}
   \centering  
   \hbox{\includegraphics[]{f1.eps}}
\caption{}
\label{fig:fig1}%
\end{figure*}

%
%
\begin{figure*}
   \centering  
   \hbox{\includegraphics[]{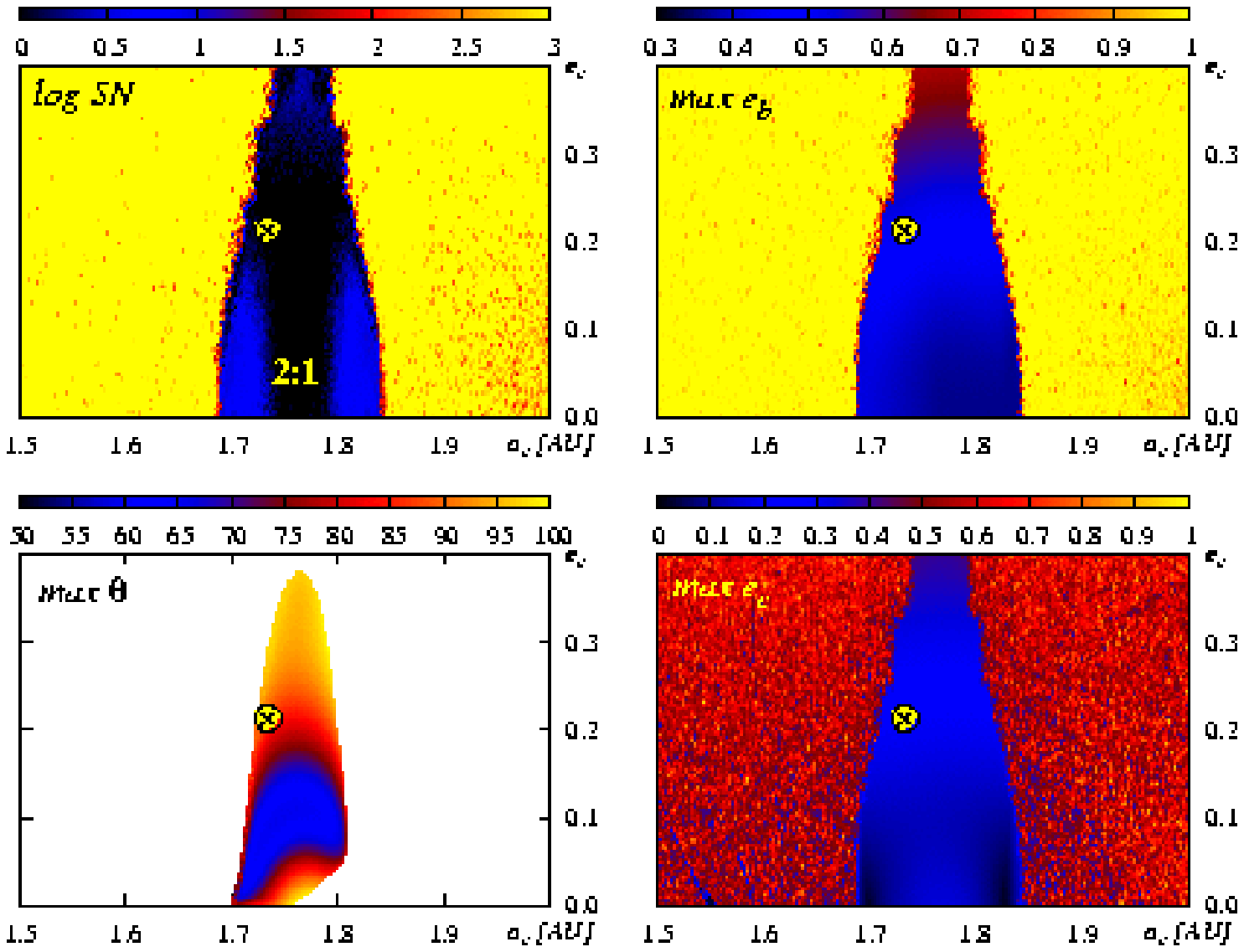}}
   \caption{}
\label{fig:fig2}%
\end{figure*}

%
%
\begin{figure*}
   \centerline{  
   \hbox{\includegraphics[]{f3.eps}}
   }   
\caption{}   
\label{fig:fig3}
\end{figure*}

%
%
\begin{figure*}
   \centerline{  
   \hbox{\includegraphics[]{f4.eps}}
   }   
\caption{}   
\label{fig:fig4}
\end{figure*}

%
%
\begin{figure*}
   \centering  
   \hbox{\includegraphics[]{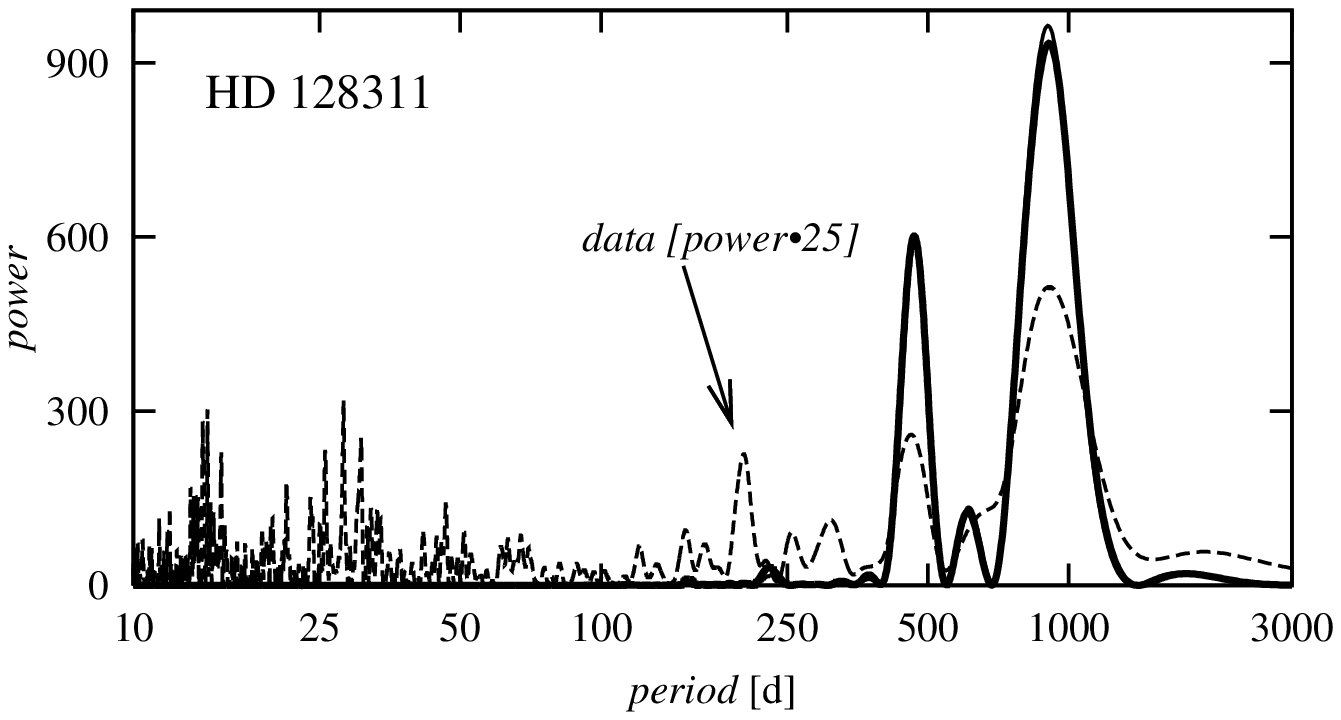}}
\caption{}   
\label{fig:fig5}%
\end{figure*}

%
%
\begin{figure*}
   \centering  
   \hbox{\includegraphics[]{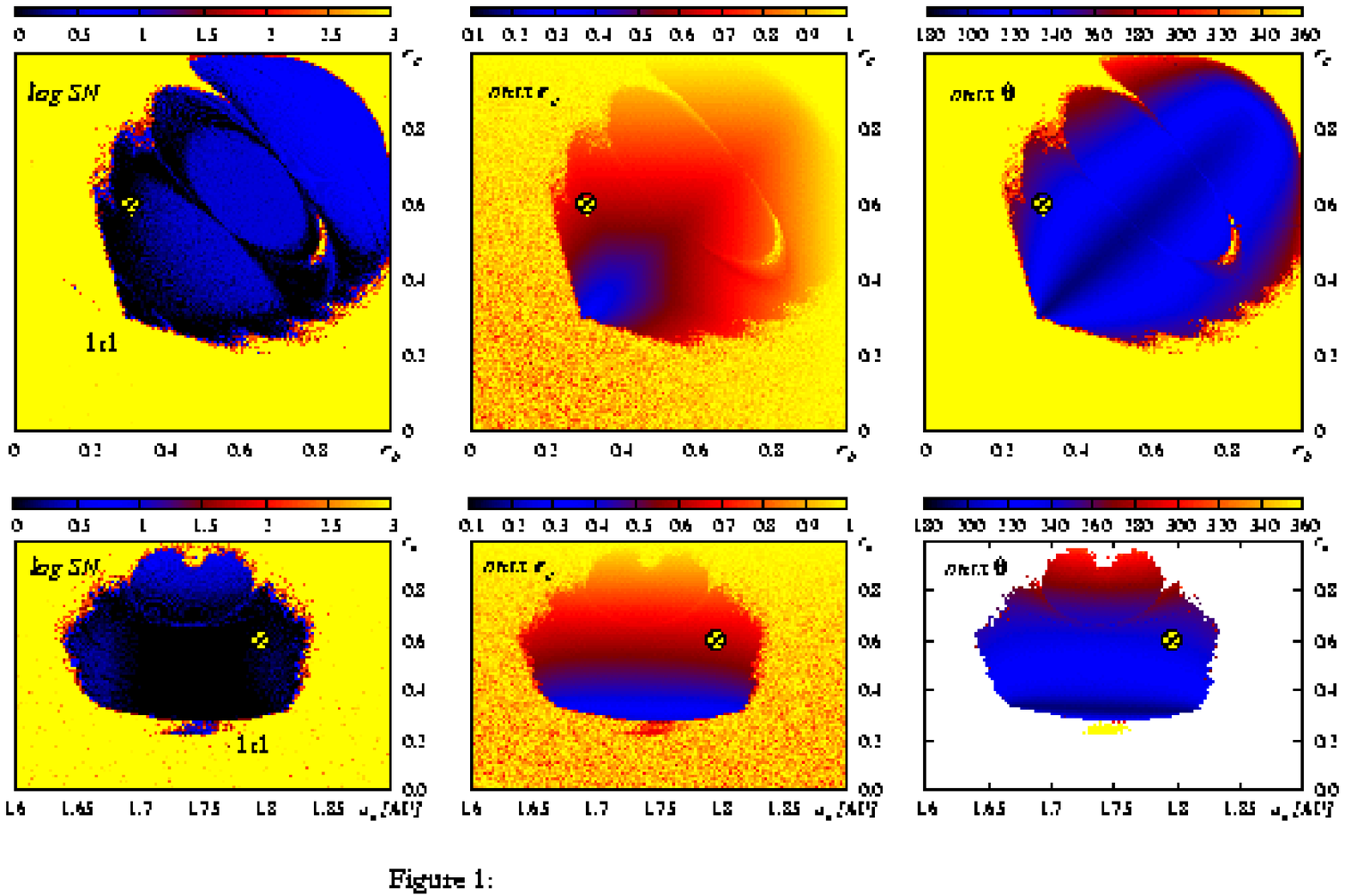}}
\caption{}   
\label{fig:fig6}%
\end{figure*}

%
%
\begin{figure*}
   \centering  
   \hbox{\includegraphics[]{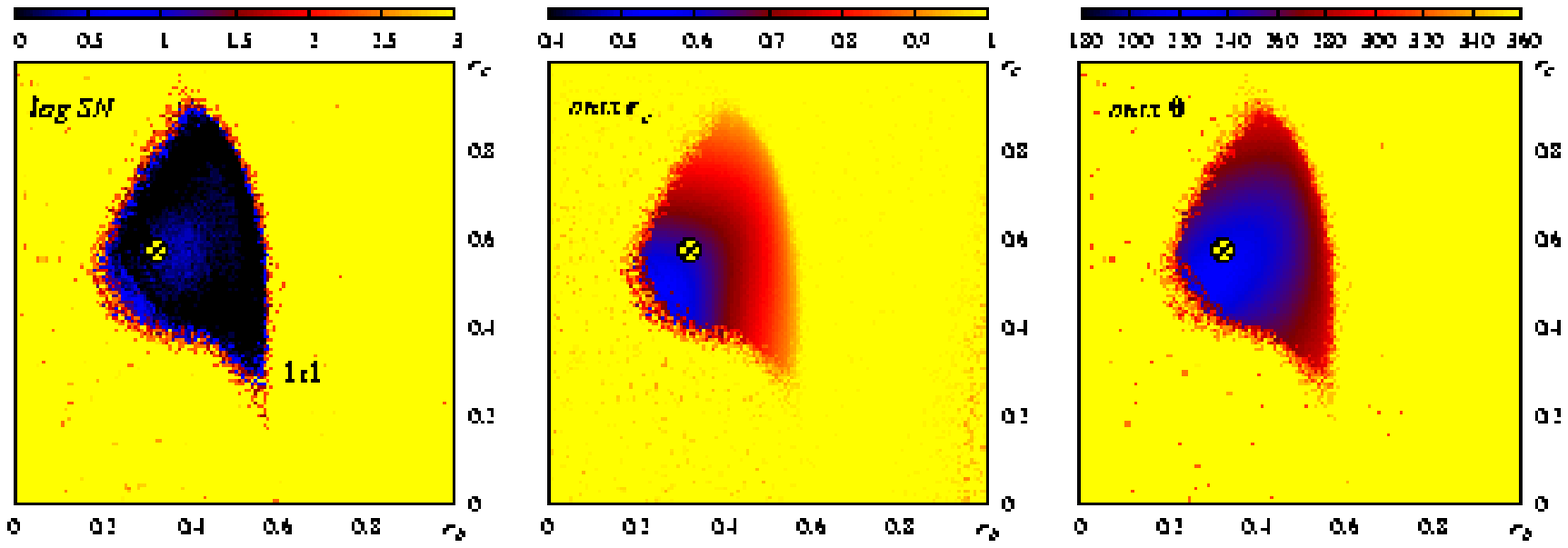}}
\caption{}   
\label{fig:fig7}%
\end{figure*}

%
%
\begin{figure*}
   \centering  
   \hbox{\includegraphics[]{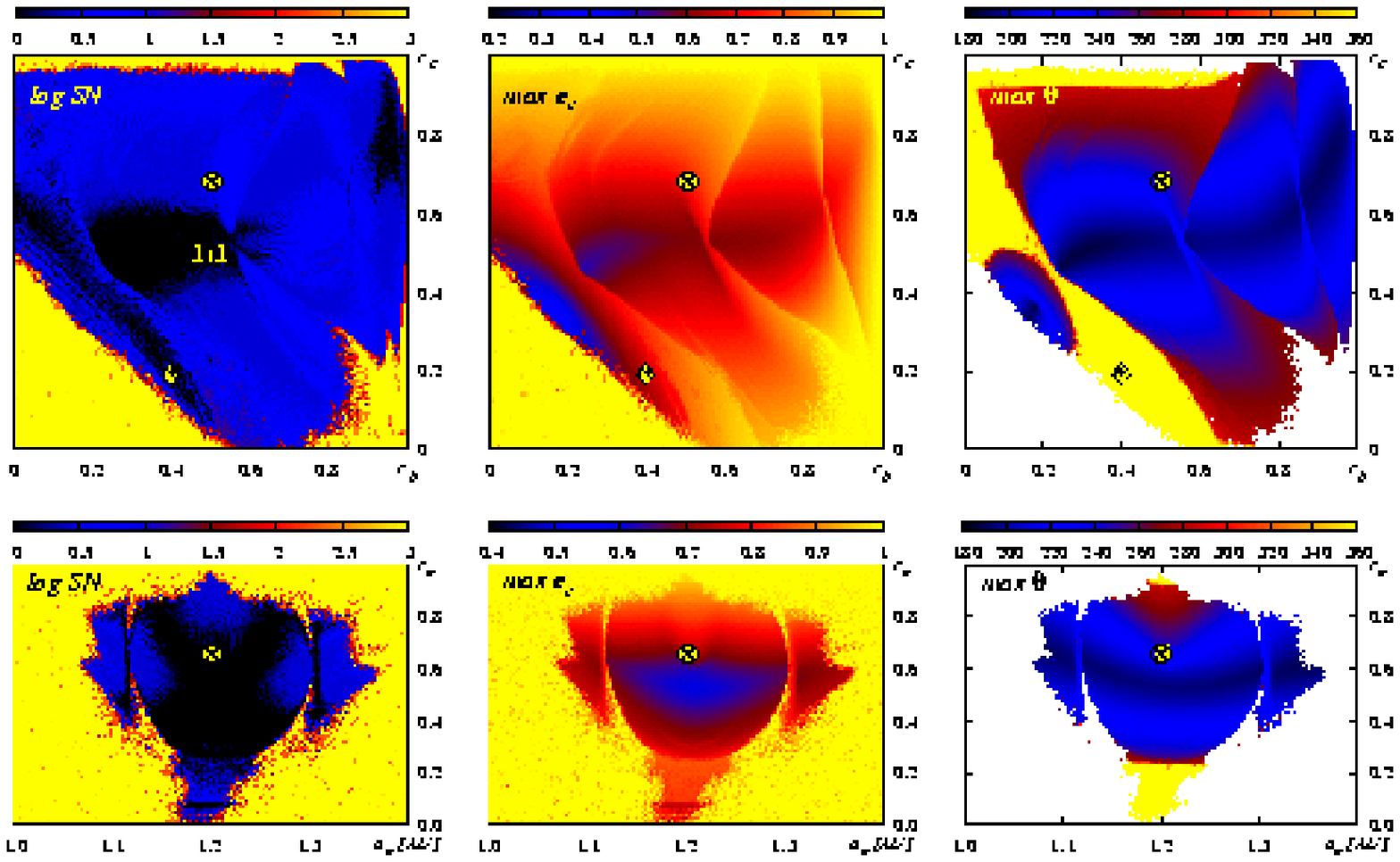}}
   \caption{}
\label{fig:fig8}%
\end{figure*}

%
%
\begin{figure*}
   \centering  
   \hbox{\includegraphics[]{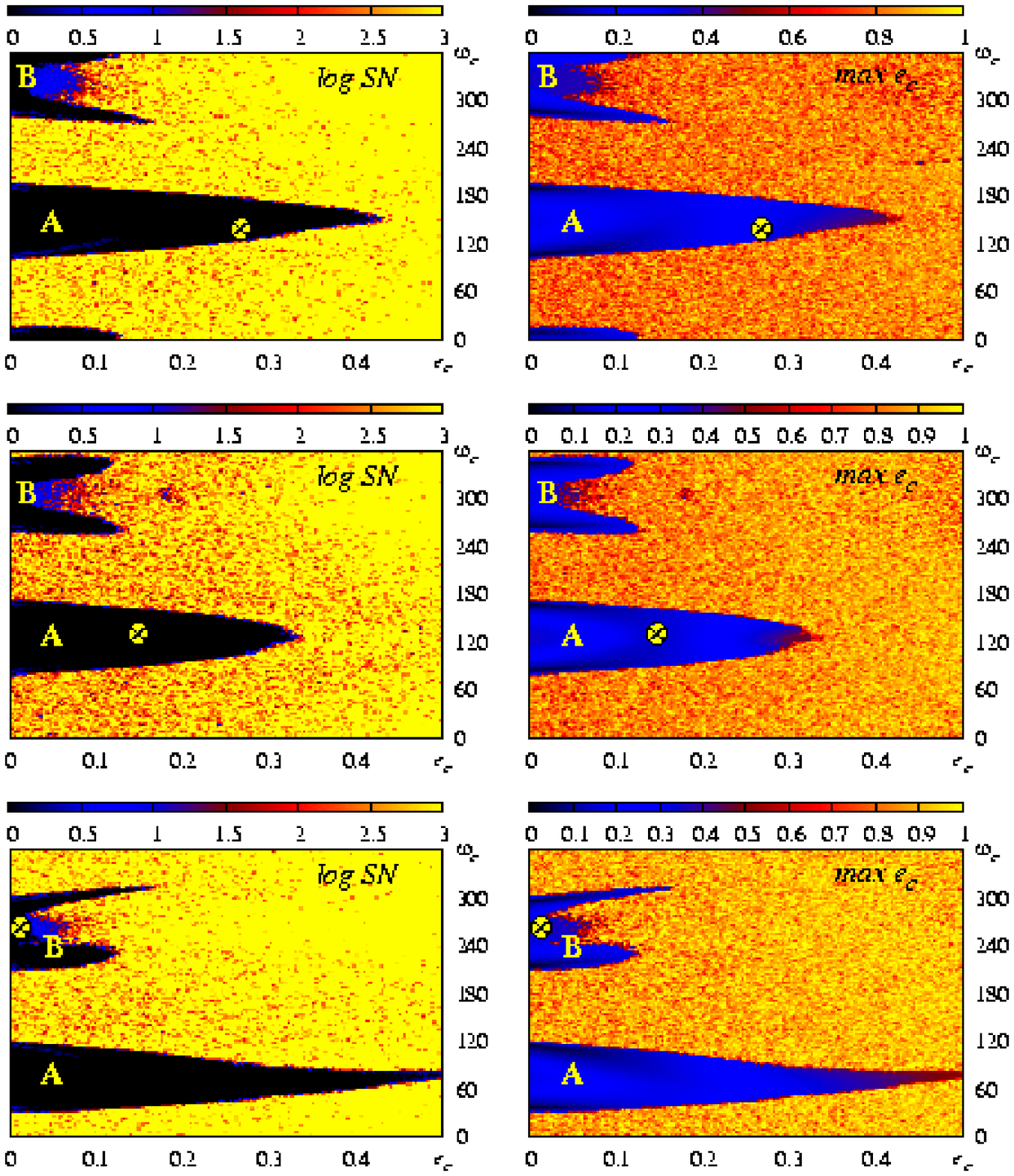}}
   \caption{}
\label{fig:fig9}%
\end{figure*}

%
%
\begin{figure*}
   \centering  
   \hbox{\includegraphics[]{f10.eps}}
   \caption{}
\label{fig:fig10}%
\end{figure*}

%
%
\begin{figure*}
   \centering  
   \hbox{\includegraphics[]{f11.eps}}
   \caption{}
\label{fig:fig11}%
\end{figure*}

%
%
\begin{figure*}
   \centering  
   \hbox{\includegraphics[]{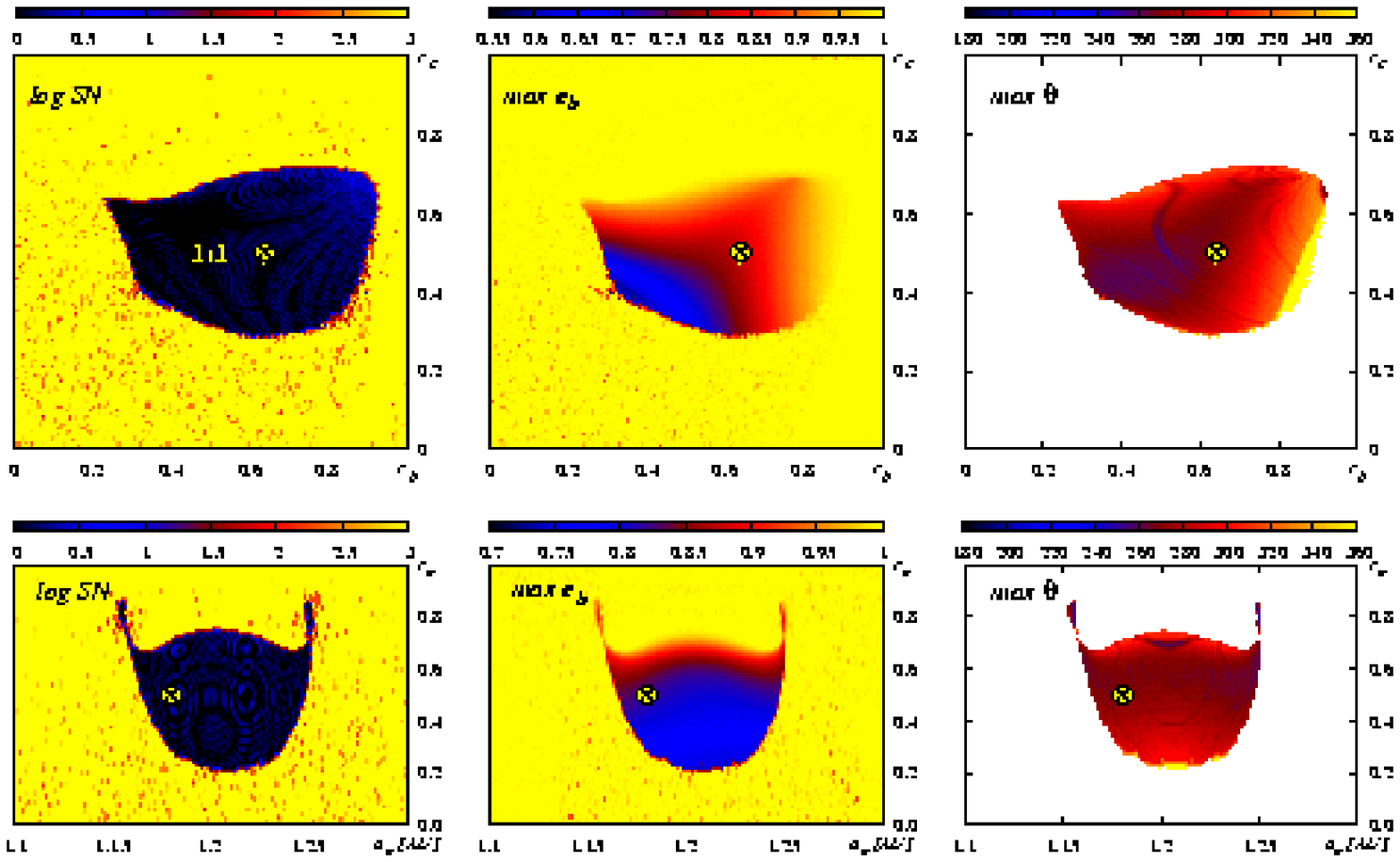}}
   \caption{}
\label{fig:fig12}%
\end{figure*}

%
%
\begin{figure*}
   \centerline{  
   \hbox{\includegraphics[]{f13.eps}}
   }   
\caption{}   
\label{fig:fig13}
\end{figure*}

%
%
\begin{figure*}
   \centering
   {  
   \hbox{\includegraphics[width=4.5in]{f14a.eps}}
   \hbox{\includegraphics[width=4.5in]{f14b.eps}}
   \hbox{\includegraphics[width=4.5in]{f14c.eps}}
   }
   \caption{}
\label{fig:fig14}%
\end{figure*}

%
%
\begin{figure*}
   \centering  
   \hbox{
   \hbox{\includegraphics[]{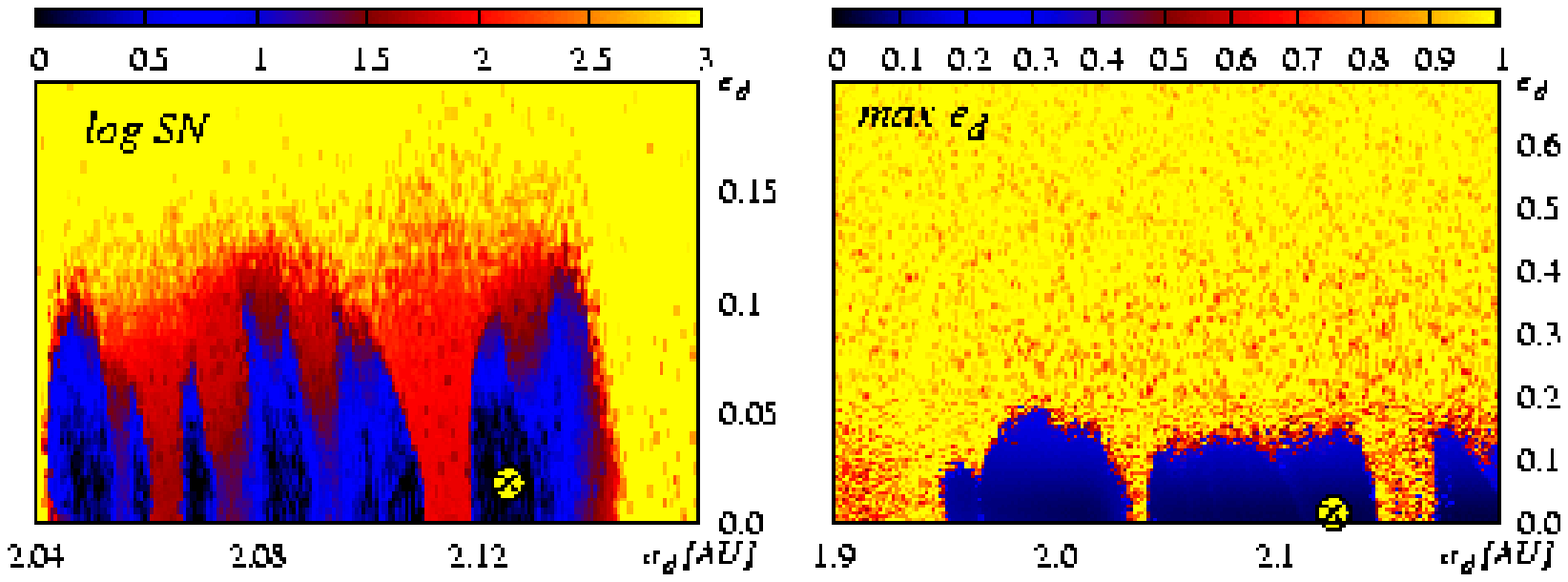}}
   }
   \caption{}

\label{fig:fig15}%
\end{figure*}

%
%
\begin{figure*}
   \centering  
   \hbox{\includegraphics[]{f16.eps}}
   \caption{}
\label{fig:fig16}%
\end{figure*}

%
%
\begin{figure*}
   \centering  
   \hbox{\includegraphics[]{f17.eps}}
   \caption{}
\label{fig:fig17}%
\end{figure*}

\end{document}